\newfont{\vs}{cmssdc10 scaled 1050}

\documentclass[useAMS,usenatbib,usegraphicx]{mn2e}
\usepackage{xspace,amsmath}
\usepackage{color,soul,epsfig,amssymb,lscape}
\usepackage{util}
\usepackage{array,colortbl}
\usepackage[dvips]{graphicx}
\usepackage{times,url,graphicx}
\usepackage{longtable}
\usepackage{txfonts}
\usepackage{natbib}
\usepackage[figuresright]{rotating}
\usepackage{subfigure}
\usepackage{multirow}
\usepackage{rotating}
\usepackage{afterpage}

\title[IFS MEGARA data of the metal-poor galaxy PHL~293B]{Mapping the ionized gas of the metal-poor HII galaxy PHL~293B with MEGARA\thanks{Based on observations collected with GTC at the Roque de los Muchachos Observatory}}
\author[C. Kehrig et al.]{C. Kehrig$^{1}$\thanks{Severo Ochoa IAA Fellow; E-mail:kehrig@iaa.es}, J. Iglesias-P\'aramo$^{1,2}$, J.M.V\'ilchez$^{1}$, A. Gil de Paz$^{3,4}$,
\newauthor
S. Duarte Puertas$^{1}$, E. P\'erez-Montero$^{1}$, A.I. D\'iaz$^{5}$, J. Gallego$^{3,4}$, E. Carrasco$^{6}$,
\newauthor
N. Cardiel$^{3,4}$, M. L. Garc\'{\i}a-Vargas$^{7}$, A. Castillo-Morales$^{3,4}$, R. Cedazo$^{8}$,
\newauthor
P. G\'omez-\'Alvarez$^{7}$, I. Mart\'{\i}nez-Delgado$^{7}$, S. Pascual$^{3,4}$, A. P\'erez-Calpena$^{7}$ \\
$^{1}$ Instituto de Astrof\'{\i}sica de Andaluc\'{\i}a, CSIC, Apartado de correos 3004, 18080 Granada, Spain \\
$^{2}$ Estaci\'on Experimental de Zonas Aridas (CSIC), Ctra. de Sacramento s/n, E-04120, Almer\', Spain \\
$^{3}$ Departamento de F\'{\i}sica de la Tierra y Astrof\'{\i}sica, Fac. CC. F{\'\i}sicas, Universidad Complutense de Madrid, Plaza de las Ciencias 1, E-28040 Madrid, Spain\\
$^{4}$ Instituto de F\'{\i}sica de Part\'{\i}culas y del Cosmos IPARCOS, Fac.CC F\'{\i}sicas, Universidad Complutense de Madrid, Plaza de las Ciencias, 1, E-28040 Madrid, Spain \\ 
$^{5}$ Departamento de F\'{\i}sica Te\'orica, Universidad Aut\'onoma de Madrid, E-28049 Madrid, Spain \\
$^{6}$ Instituto Nacional de Astrof\'{\i}sica, \'Optica y Electr\'onica, INAOE, Calle Luis Enrique Erro No.1, C.P. 72840, Tonantzintla, Puebla, Mexico  \\
$^{7}$ FRACTAL, S.L.N.E., Calle Tulip{\'a}n 2, portal 13, 1A, E-28231 Las Rozas de Madrid, Spain \\
$^{8}$ Universidad Polit\'ecnica de Madrid, 28031, Madrid, Spain}

\begin{document}

\date{Accepted Date. Received Date; in original Date}

\pagerange{\pageref{firstpage}--\pageref{lastpage}} \pubyear{2016}

\maketitle

\label{firstpage}

\begin{abstract}
Here we report the first spatially resolved spectroscopic study for
the galaxy PHL~293B using the high-resolution GTC/MEGARA integral
field unit (IFU). PHL~293B is a local, extremely metal-poor, high
ionization galaxy. This makes PHL~293B an excellent analogue for
galaxies in the early Universe. The MEGARA aperture ($\sim$ 12.5''
$\times$ 11.3'') covers the entire PHL~293B main body and its far-reaching ionized gas. We created and discussed maps of all
relevant emission lines, line ratios and physical-chemical properties
of the ionized ISM. The narrow emission gas appears to be ionized
mainly by massive stars according to the observed diganostic line
ratios, regardless of the position across the MEGARA aperture. We
detected low intensity broad emission components and blueshifted
absorptions in the Balmer lines (H$\alpha$,H$\beta$) which are located
in the brightest zone of the galaxy ISM. A chemically homogeneity,
across hundreds of parsecs, is observed in O/H. We take the oxygen
abundance \mbox{12+log (O/H) = 7.64 $\pm$ 0.06} derived from the
PHL~293B integrated spectrum as the representative metallicity for the
galaxy. Our IFU data reveal for the first time that the nebular
HeII$\lambda$4686 emission from PHL~293B is spatially extended and
coincident with the ionizing stellar cluster, and allow us to compute
its absolute HeII ionizing photon flux. Wolf-Rayet bumps are not detected
excluding therefore Wolf-Rayet stars as the main HeII excitation
source. The origin of the nebular HeII$\lambda$4686 is discussed.

\end{abstract}
%

\section{Introduction}

HII galaxies are the most metal-poor starbursts in the local
Universe \citep[e.g.,][]{W04,K06,I12,J17}. These galaxies present
intense star-formation rates, and they usually have low masses and blue
optical colours. The hot, luminous massive stars present in HII
galaxies give off vast quantities of high-energy UV photons which
ionize the gas producing strong nebular emission-line
spectra \citep[e.g.,][]{K04,C09mrk1418,C10}.

\begin{figure}
\centering
\includegraphics[width=7.5cm]{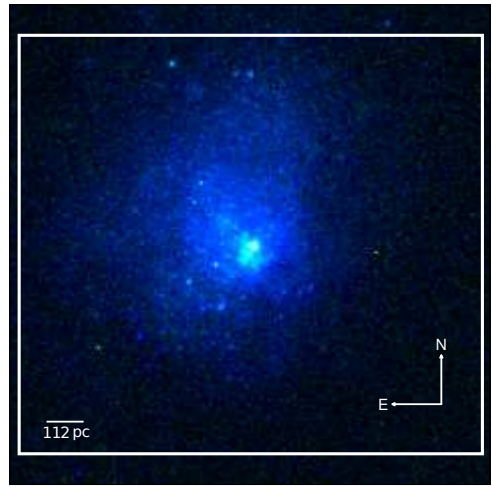}
\caption{Colour composite HST image of PHL~293B generated using the
  following bands: blue = WFC3/F606W, green = WFC3/438W, red = WFC3/F814W (HST Proposal ID 12018; PI: A.Prestwich). The white box denotes the observed field of view (FOV) of MEGARA (12.5'' $\times$ 11.3''). North is up and east is to the left.}
\label{megara_ifu}
\end{figure}

PHL~293B is a very compact HII galaxy \citep[effective radius of its star-forming component $\sim$ 0.7''; e.g.,][]{P08} which belongs to the ``Palomar-Haro-Luyten'' survey of faint
 galaxies \citep[see][]{F80,K81}. The ionized gas of
   PHL~293B presents a very low oxygen abundance of 12+log(O/H) $\approx$ 7.6-7.7 [$\sim$ 1/10 solar metallicity\footnote{assuming a solar abundance 12+log(O/H)$_{\odot}$ = 8.69 \citep{asplund09}}; e.g., \citet{F80,K81,P08,I11,F18}].  Moreover, PHL~293B shows ultra-high excited gas indicated by the
presence of the nebular HeII$\lambda$4686 emission
\citep[e.g.,][]{F80,I07}, and its very high specific star formation
rate \citep[sSFR = SFR/M$_{\star}$ $\sim$ 6 Gyr$^{-1}$; see table 4
from][]{F13} is comparable to those found in the high-redshift
Universe \citep[e.g.,][]{DS16}. These features are more 
  commonly observed and predicted in distant star-forming galaxies in
  comparison with local starbusrst \citep[e.g.,][]{len15,mainali18,izotov19,sobral19}. This
  makes PHL~293B a remarkable place nearby that allows us to
  study in detail physical conditions which may be predominant in primeval starbursts
 \citep[see also][]{K16,guseva17,izotov18,K18,S19}. The
optical spectrum of PHL~293B shows the typical strong narrow emission
lines normally seen in the spectra of HII galaxies. Besides, its spectrum exhibits other features as low-intensity broad wings and blueshifted narrow absorptions in the hydrogen recombination lines \citep[e.g.,][and references therein]{IT09,T14}.
Table~\ref{sample} lists other general properties of PHL~293B.
Fig.~\ref{megara_ifu} shows a three-colour composite image of
PHL~293B from the Hubble Space Telescope (HST)/WFC3 which reveals an
extended gaseous nebulae and star-forming activity mainly present in the southern zone of the galaxy \citep[see also][]{P08}. 

In the last years, spatially resolved spectroscopy has opened a new
window onto our understanding of the ionized gas in low-redshift SF galaxies,
preventing us from an over-simplified view of
it \citep[e.g.,][]{K08,K13,salva19,u19,sanchez20}. In this work, we present the
first 2D spectroscopic study of PHL~293B based on commissioning observations with
the {\it Multi Espectr\'{o}grafo en GTC de Alta Resoluci\'{o}n para
Astronom\'{\i}a} (MEGARA; see next section).  Our MEGARA data provide a
detailed scanning of the structure and properties of the PHL~293B
ionized gas. Moreover, we
derive the first integrated spectrum and total HeII-ionizing photon
flux from PHL~293B.

The paper is organized as follows. In Section 2, we report
observations and data reduction. Flux measurements and emission line
intensity maps are presented in Section 3. In Section 4 we show the
2D view of the ionization structure and nebular properties.  In
Section 5, we present the integrated properties from selected regions
of PHL~293B. Section 6 discusses the spatially resolved 
HeII$\lambda$4686-emitting region and the origin of the HeII
excitation. Finally, Section 7 summarizes the main conclusions derived
from this work.

\begin{table}
\caption{General Properties of PHL~293B\label{sample}}
\centering
\begin{tabular}{lc} \hline
Parameter      & PHL~293B\\ \hline
Alternate names  & Kinman Dwarf, A2228-00        \\
R.A. (J2000.0)  & 22h 30m 36.8s         \\
DEC. (J2000.0)      & -00$^{\circ}$ 06' 37''        \\
redshift   & 0.0051               \\
D$^{a}$(Mpc) & 23.1   \\
HType$^{b}$ & Im? \\
Scale (pc/$\prime\prime$)   & 112  \\
(B-V)$^c$ & 0.56  $\pm$ 0.05         \\
(V-R)$^c$ & 0.27  $\pm$ 0.04         \\
M$_{B}$$^c$(mag) & -14.37    \\
A$_{V}$$^d$(mag) & 0.193   \\
\hline 
\end{tabular}
\begin{flushleft}
$^{a}$ Distance taken from the NASA/IPAC Extragalactic Database (NED); $^{b}$ Hubble Type from NED; $^{c}$ From \cite{C01}; $^{d}$ Galactic extinction from \cite{s11} 
\end{flushleft}
\end{table}

\section{Observations and data reduction}

The data of PHL~293B were obtained with MEGARA \citep[see][]{gilspie18,carspie18}, attached to
the 10.4m GTC telescope at the Roque de los Muchachos
Observatory. Observations were taken during the second commissioning
run in 2017 July 25$^{th}$ and 29$^{th}$, using the Large Compact Bundle (LCB)
IFU mode which provides a field of view (FOV) of
12.5$\times$11.3~arcsec$^{2}$ ($\sim$ 1.4 kpc $\times$ 1.3 kpc at the
distance of 23.1 Mpc; see Fig.~\ref{megara_ifu}), with a
spaxel\footnote{Individual elements of IFUs are usually named
``spatial pixels'' (so-called ``spaxel''); the term is
used to distinguish a spatial element on the IFU from a detector pixel.}
diameter of 0.62~arcsec. In order to cover the main optical
emission-lines, the observations were carried out with three gratings;
we used the ``blue'' VPH405-LR (centered at 4025 \AA) and ``green'' VPH480-LB (centered at 4785 \AA) gratings which give spectral ranges ($\AA$)/dispersions ($\AA$~pix$^{-1}$) of $\sim$ 3653-4386/0.17 and $\sim$ 4332-5196/0.20, respectively. On the
red side, the VPH665-HR (centered at 6602 \AA) was utilized, providing a
spectral range from $\sim$ 6445-6837 and 0.09 $\AA$/pix. The resolving power of the gratings are $\sim$ 6000 in case of LR VPHs and $\sim$ 20000 for VPH665-HR.

We observed a total of 2.25 hours on the galaxy, with the integration
time split into three exposures for each VPH: $3\times 600$~s for
VPH665-HR and VPH480-LR each, and $3\times 1500$~s for VPH405-LR.  The
seeing was about 1~arcsec and 0.5~arcsec during the first and second
observing nights, respectively. All science frames were observed at
airmasses $\lesssim$ 1.2 to minimize the effects due to differential
atmospheric refraction. Additionally, all necessary calibration frames
(exposures of arc lamps and of continuum lamps) were obtained.

The data reduction and sky subtraction were carried out using the
MEGARA Pipeline as described in
\cite{p19}. Due to severe haze throughout the observing run, we did not flux-calibrate the data using standard stars.
Instead, we flux-calibrated our spectra using the Sloan Digital Sky
Survey (SDSS) spectrum of PHL~293B. First, we co-added the spaxels within a
3~arcsec-diameter aperture (i.e., the same size of the SDSS fiber),
centered at the brightest spaxel of the LCB IFU, to create a 1D spectrum called the
``SDSS-like'' MEGARA spectrum. We then measured the flux
of the emission lines present in both the SDSS and SDSS-like
MEGARA spectra. We compared these flux measurements by performing a third
order logarithm polynomial fit to obtain the sensitivity function which was applied to the science frames.
Given that the SDSS spectrum (spectral range $\sim$ 3800-9000 \AA) of PHL~293B does not cover our bluest emission lines [O{\sc ii}]$\lambda\lambda$3727,3729\AA, the sensitivity function was extrapolated to allow for their relative flux calibration.    

\section{Flux measurements and emission line intensity maps}\label{flux_measurements}

Here we measure emission line fluxes from individual spaxels
based on our own IDL scripts. On top of a linear flat continuum, we
fit a Gaussian profile to each emission line using the IDL based
routine MPFIT \citep[][]{ma09}; the peak intensity, the line width
$\sigma$ and the central wavelength $\lambda_{C}$ for each line are
kept as free parameters. Note that, due to the high spectral
resolution of MEGARA we were able to resolve the [OII] doublet and,
consequently, measure its individual lines at $\lambda$3726~\AA\ and
$\lambda$3729~\AA. In the case of the H$\alpha$+[N{\sc ii}] lines, we
perform a simultaneous fit keeping a nitrogen [N{\sc
ii}]$\lambda$6584/[N{\sc ii}]$\lambda$6548 line ratio of 3. Previous
work, based on single-aperture/long-slit spectroscopy of the central
star-forming (SF) knot of PHL~293B, have detected the presence of
several components for the H$\alpha$, H$\beta$
lines \citep[e.g.,][]{IT09,T14,F18}. Following these authors, we fit
these Balmer lines assuming three Gaussian components: narrow + broad
emission, and one absorption component.  Errors in the derived
parameters (line flux, peak intensity, line width $\sigma$, central
wavelength $\lambda_{C}$) are estimated by using the bootstrap method.

\begin{figure*}
\center
\includegraphics[bb=4 4 793 648,width=0.43\textwidth,clip]{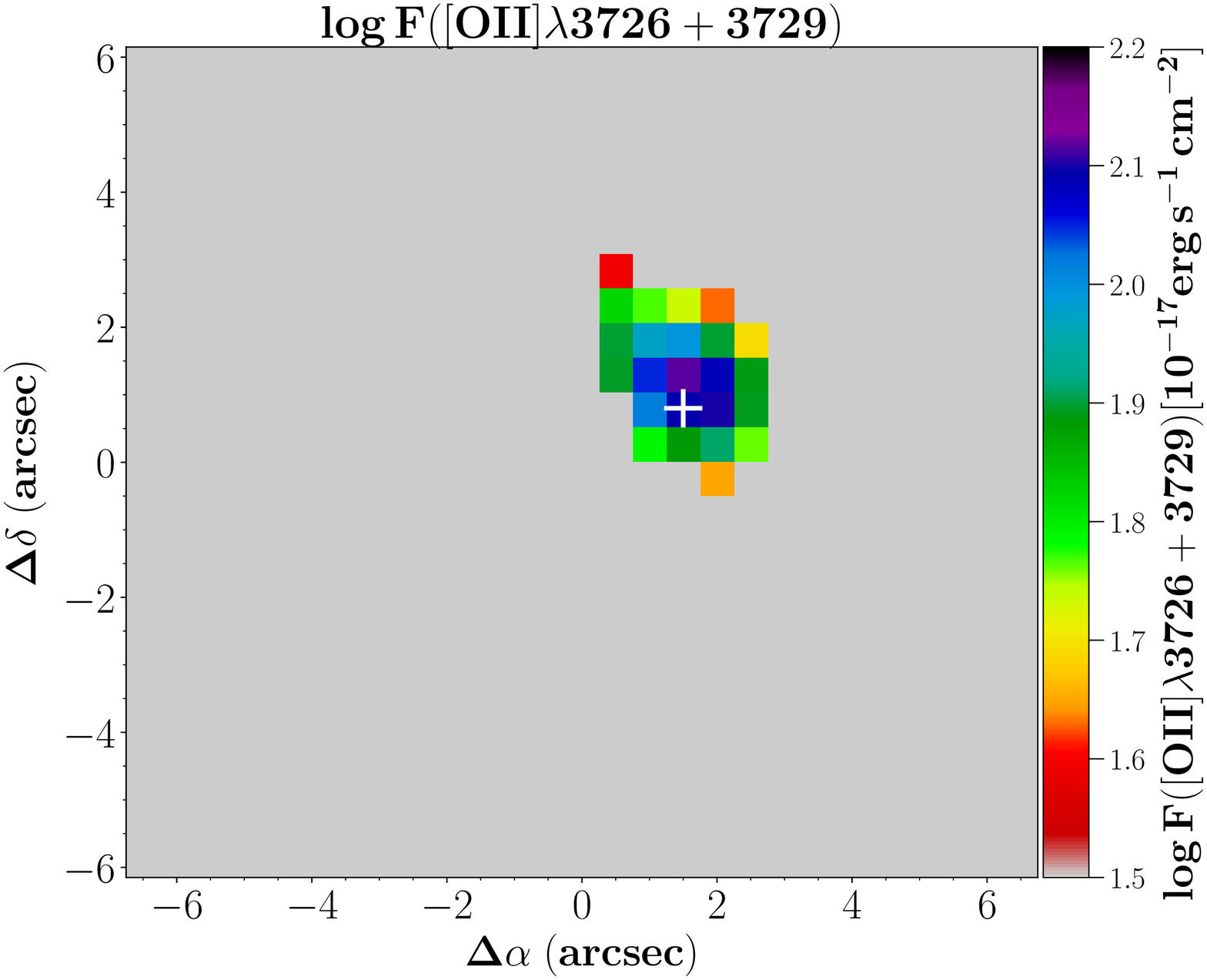}
\includegraphics[bb=4 4 793 648,width=0.43\textwidth,clip]{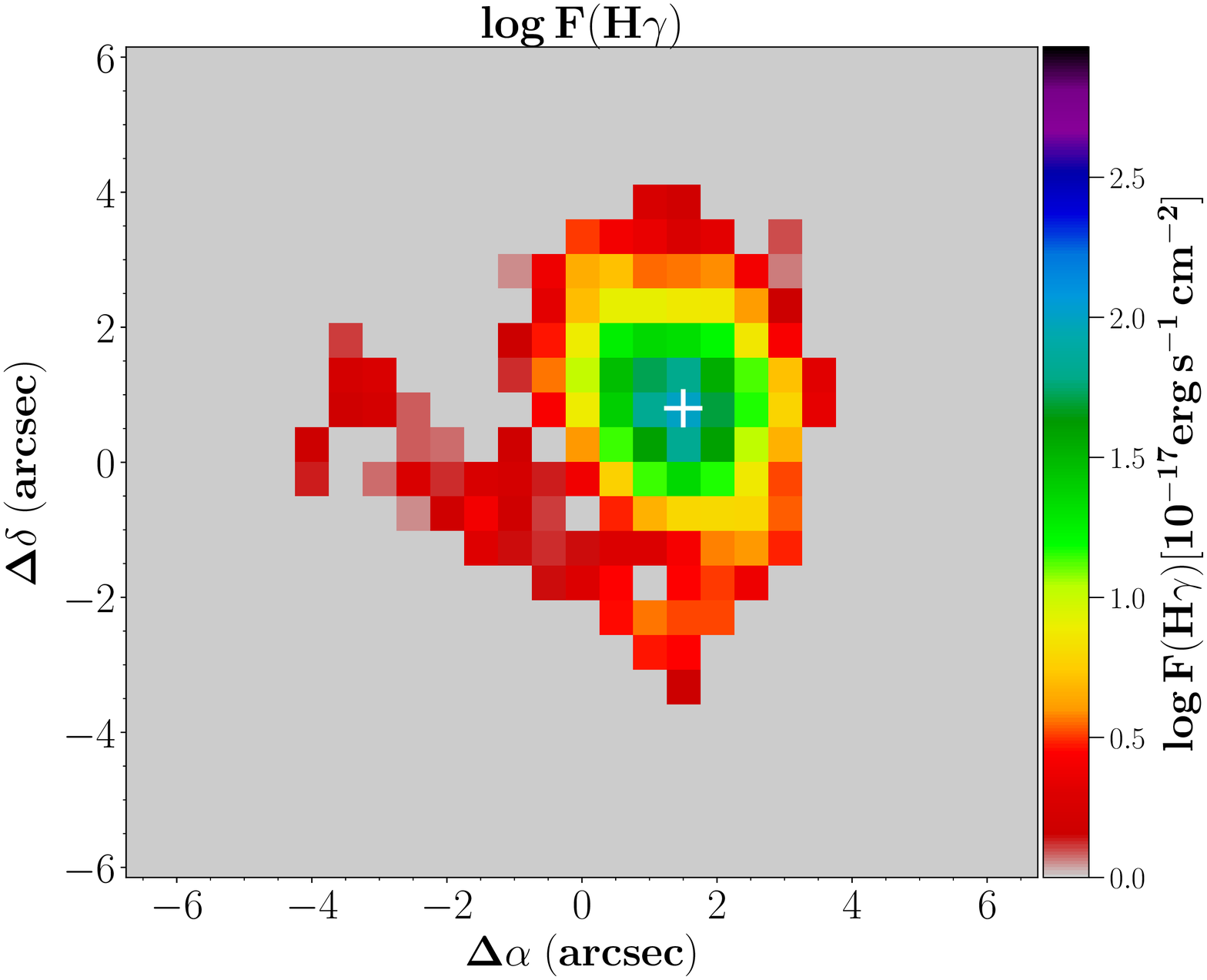}\\
\includegraphics[bb=4 4 793 648,width=0.43\textwidth,clip]{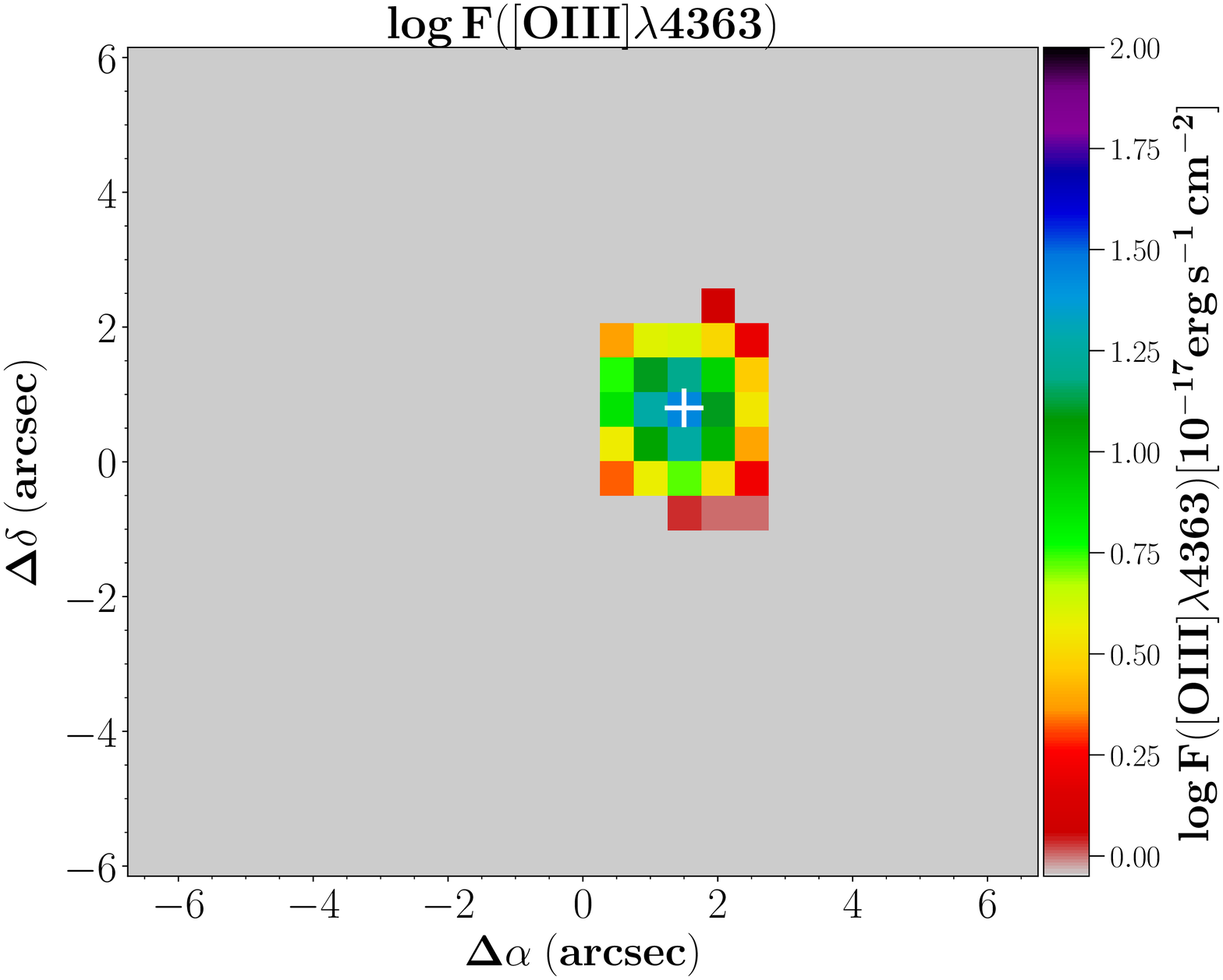}
\includegraphics[bb=4 4 793 648,width=0.43\textwidth,clip]{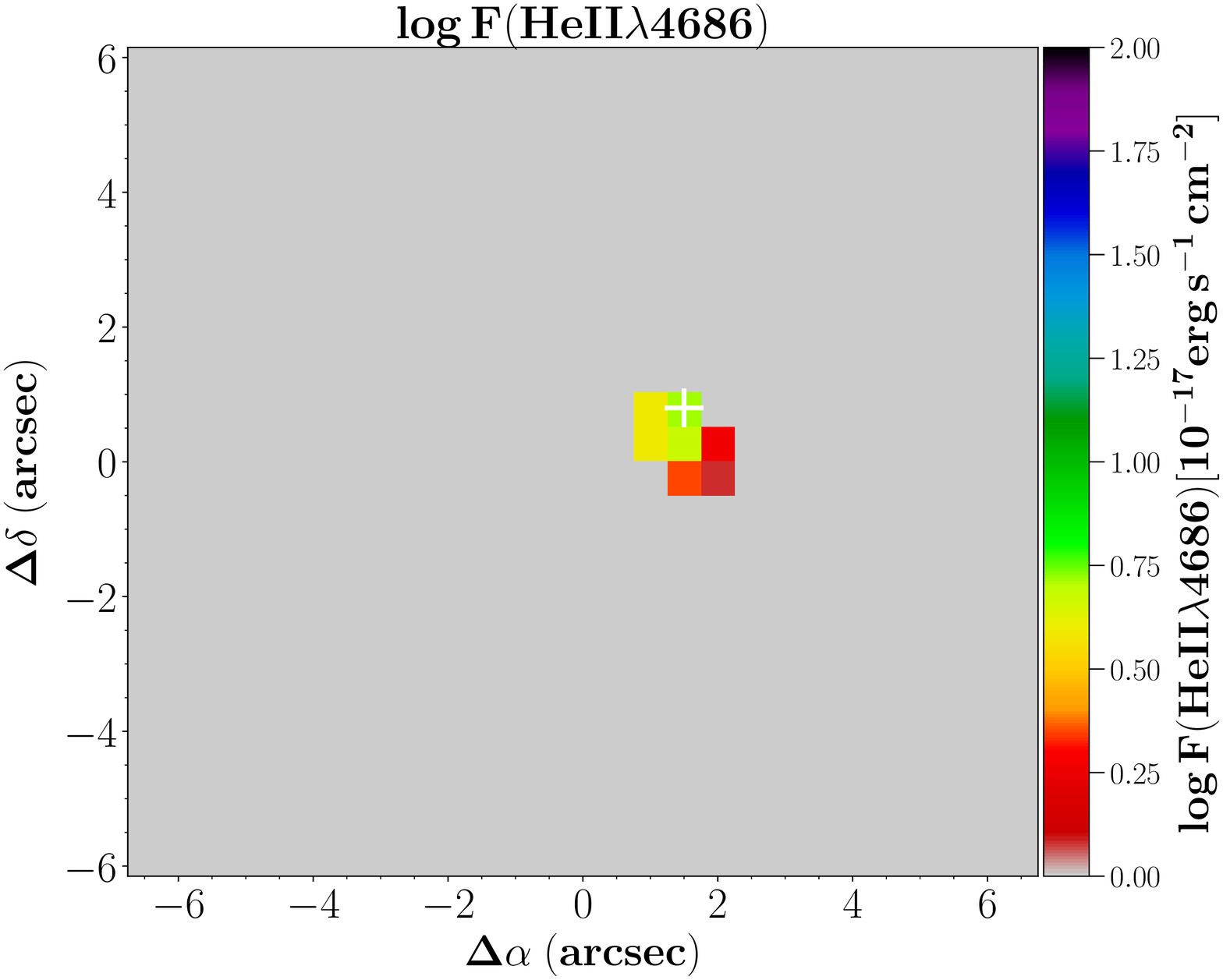}\\
\includegraphics[bb=4 4 793 648,width=0.43\textwidth,clip]{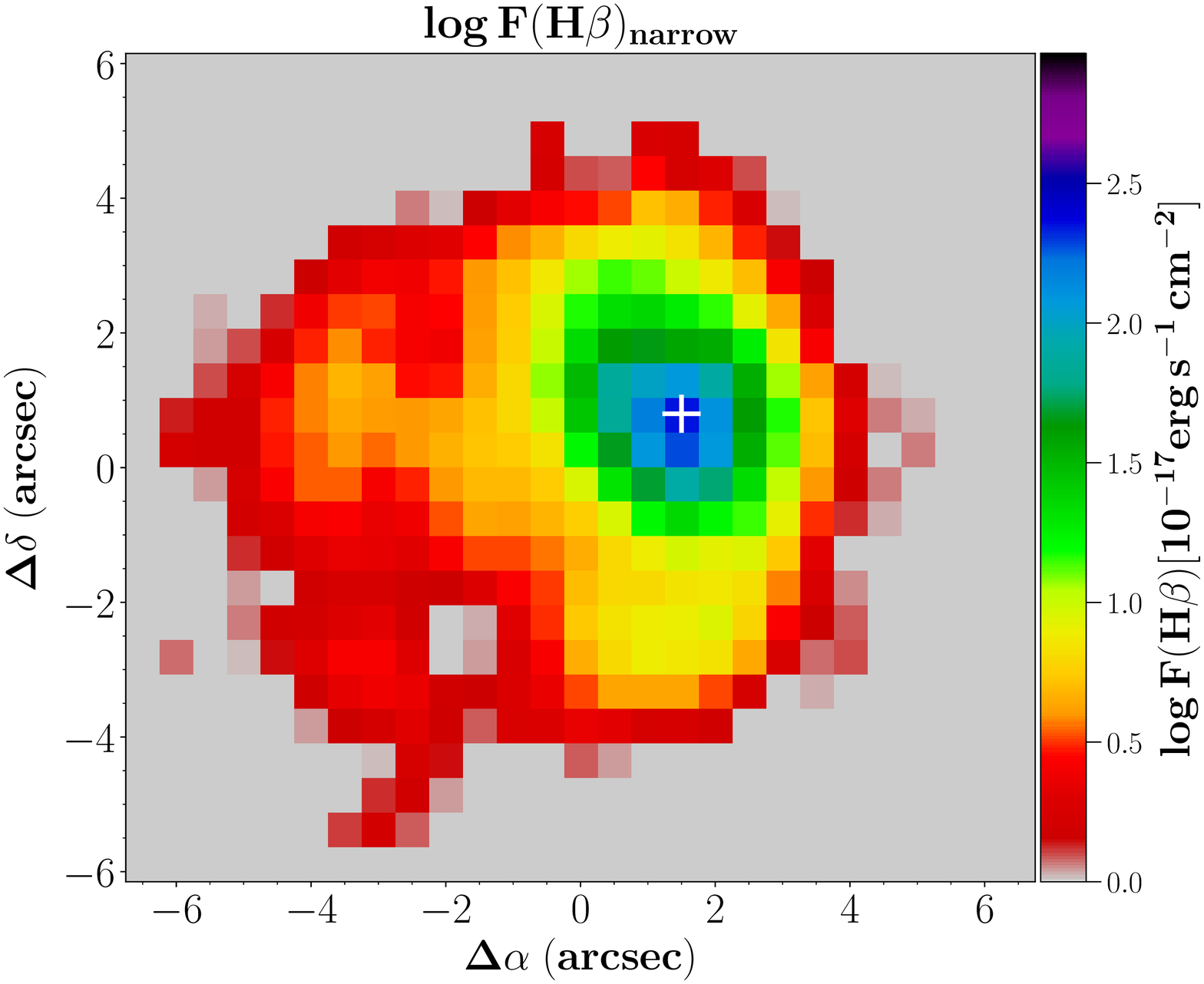}
\includegraphics[bb=4 4 793 648,width=0.43\textwidth,clip]{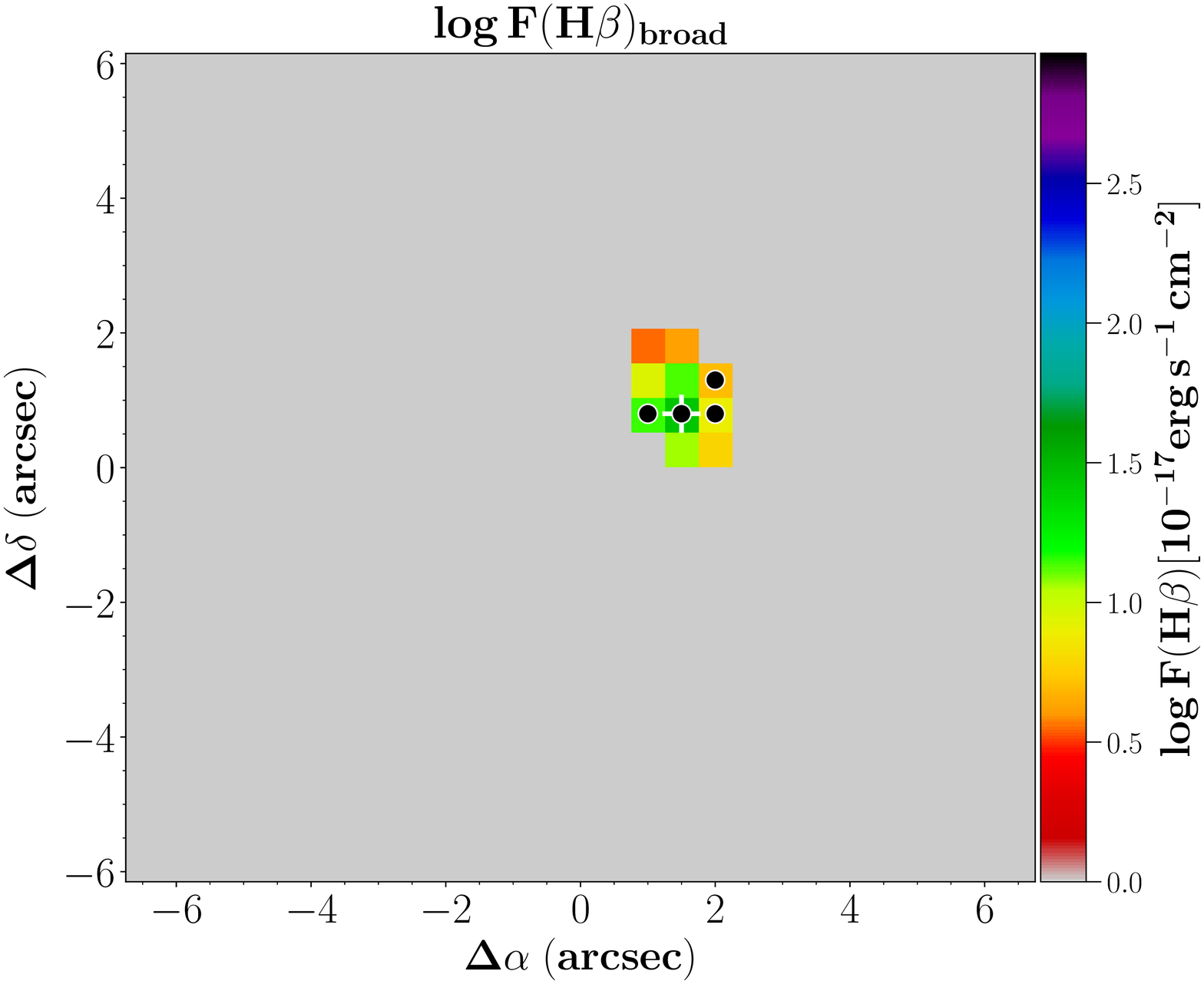}\\
\caption{Flux maps in logarithmic scale. Only fluxes with
    S/N $>$ 3 are shown. East is left and North is up. The plus (+) sign indicating the H$\alpha$ emission peak is plotted for orientation. The spaxels with no measurements available are left grey. The spaxels where we detect a P Cygni-like profile in the H$\beta$ line are indicated with a black circle on the map for the H$\beta$ broad emission.} 
\label{line_maps} 
\end{figure*}

By combining the line fluxes with the position of the spaxels on the
sky, we create all maps presented in this
paper. Figure~\ref{line_maps} exposes the intensity maps for several
emission lines; only fluxes with S/N $>$ 3 are
displayed.\footnote{Here the S/N is defined as the ratio between the
peak intensity of the emission line and the standard deviation of its underlying continnum.}  We show for the first time
the spatial distribution for the broad H$\alpha$,H$\beta$ components
for PHL~293B.  The global spatial structure of the brightest lines
(narrow H$\beta$, [O{\sc iii}]$\lambda$5007, and narrow H$\alpha$) is
similar, with [O{\sc iii}]$\lambda$5007 and narrow H$\alpha$ emission
covering almost the entire FOV. The spatial distribution of the
fainter lines (e.g., [O{\sc iii}]$\lambda$4363, HeII$\lambda$4686,
[N{\sc ii}]$\lambda$6584, [S{\sc ii}]$\lambda$6717+6731), and the
broad H$\beta$ and H$\alpha$ are restricted to the inner parts of the
galaxy.

When comparing the [O{\sc ii}] emission distribution to that for
other relatively bright lines as H$\beta$, we find the former to be
more compact. This could be related partially to the fact
that the [O{\sc ii}] lines lie at the blue edge of the spectra
(i.e. $\lambda$ $<$ 3750 \AA) where we not only observe lower S/N but
also expect less accuracy of the flux calibration \citep[see Section 2
for details; see e.g.,][]{sanchez12,yan2016,lopez2019}. However, we
highlight that the PHL~293B ionization structure which seems to be
dominated by high excitation should also plays an important role; e.g.,
the [O{\sc iii}]$\lambda$5007 emission is spatially
wide-ranging as long as the [O{\sc ii}]$\lambda$$\lambda$3726,3729, [N{\sc ii}]$\lambda$6584 and [S{\sc ii}]$\lambda$$\lambda$6717,6731 lines
extend over a much smaller area (see Fig.~\ref{line_maps}). The
likeness between the maps of [O{\sc ii}], [N{\sc ii}] and [S{\sc ii}]
is generally expected due to their similar ionization potential, albeit [N{\sc
ii}] and [S{\sc ii}] lines sit at the red part of the spectra where S/N and
flux calibration effects mentioned above should be minor.

The emission of both, the high and low intensity lines, are peaked
on the southern HII region where the star formation is
mostly concentrated \citep[the bright blue knot in Fig.~\ref{megara_ifu}; see also][]{P08}.
In agreement with previous work \citep[e.g.,][]{IT09,T14} we were able
to detect H$\beta$ and H$\alpha$ P Cygni-like profiles in some
integrated spectra (see Section~\ref{int} for details) and in a few
individual spaxels which are indicated in the map of the H$\beta$ and
H$\alpha$ broad emission components (see
Fig.~\ref{line_maps}). P Cygni-like profiles are displayed in
Figs.~\ref{pcyg_hb} and \ref{pcyg_ha}.

\addtocounter{figure}{-1}

\begin{figure*}
\center
\includegraphics[bb=4 4 793 648,width=0.43\textwidth,clip]{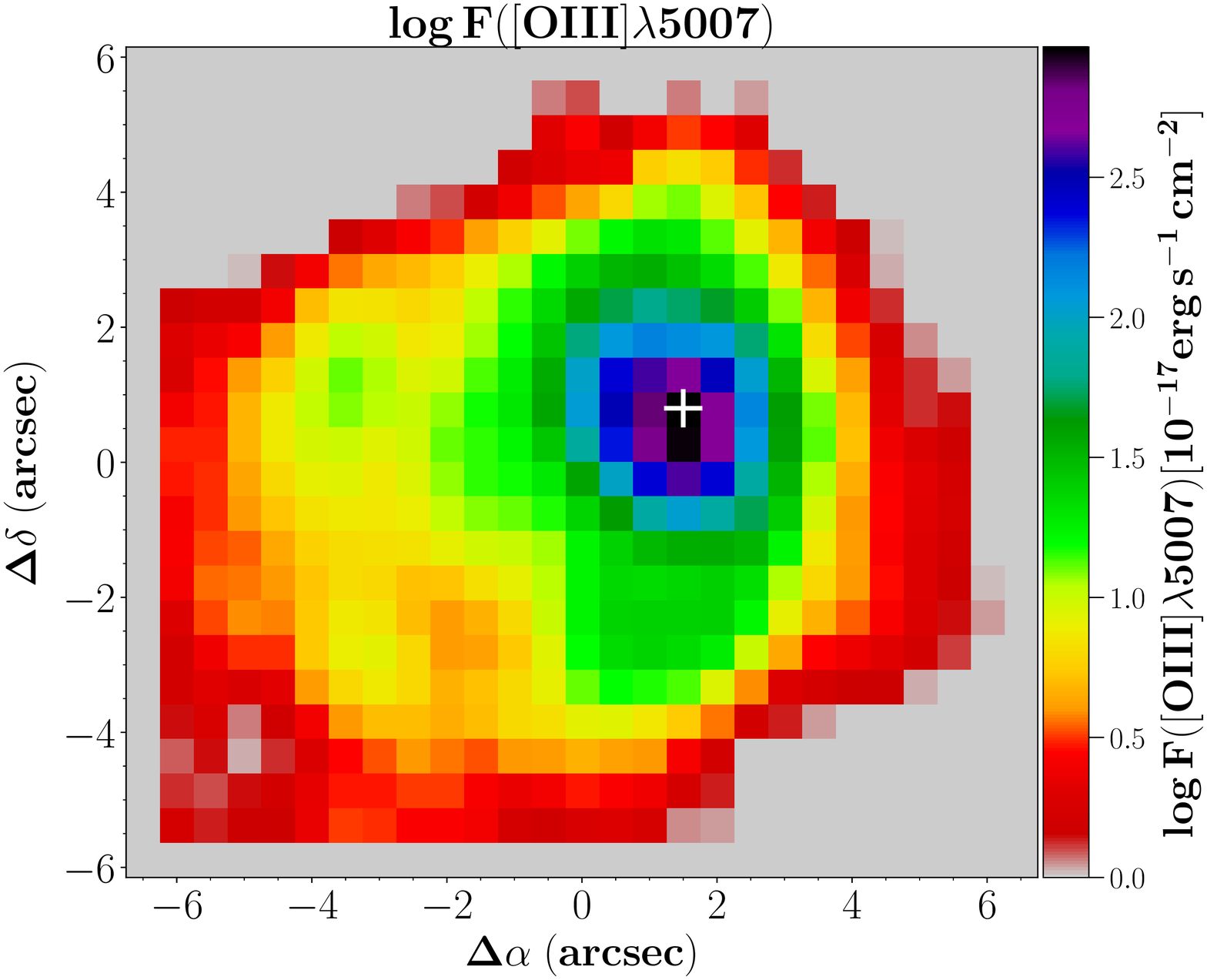}
\includegraphics[bb=4 4 793 648,width=0.43\textwidth,clip]{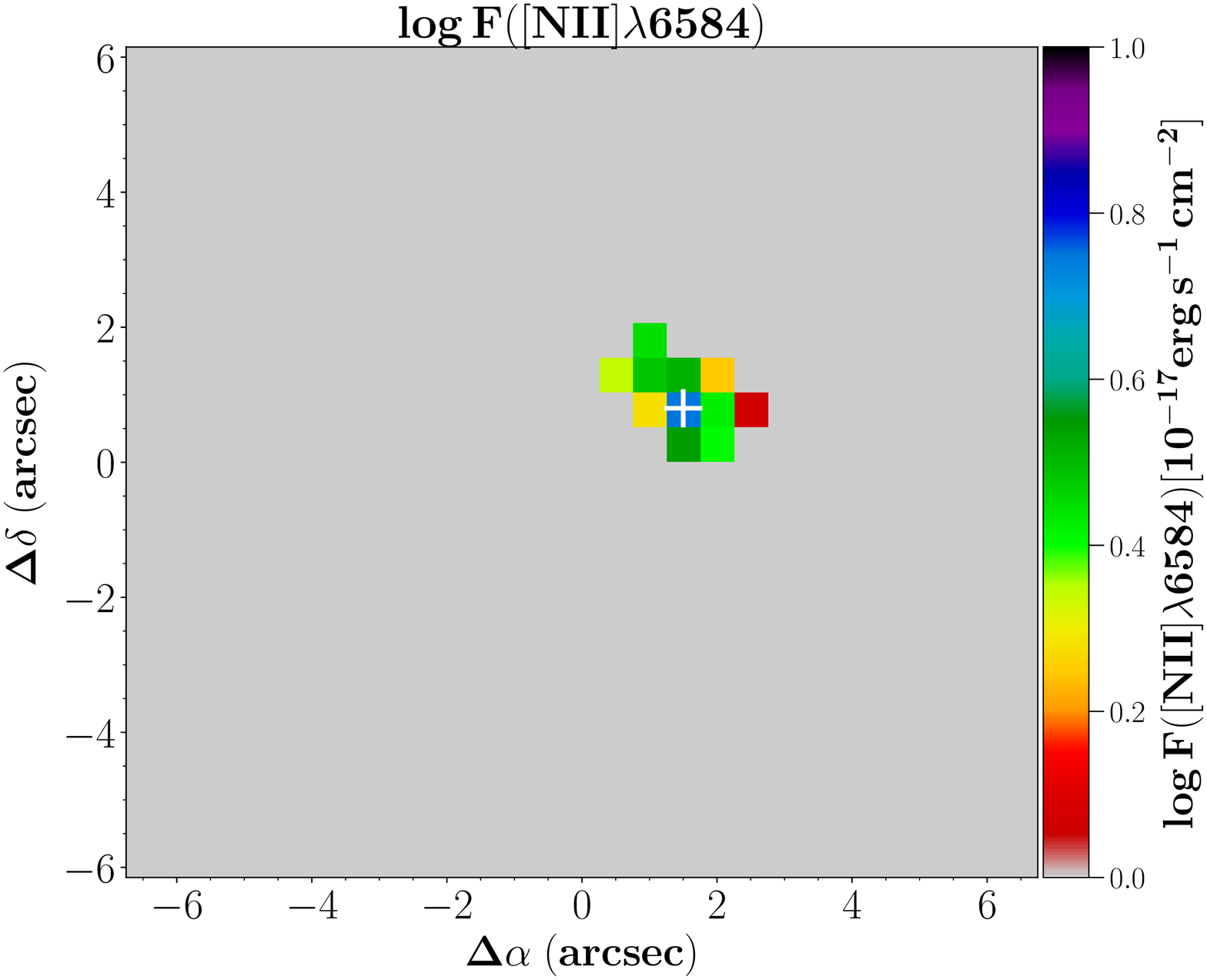}\\
\includegraphics[bb=4 4 793 648,width=0.43\textwidth,clip]{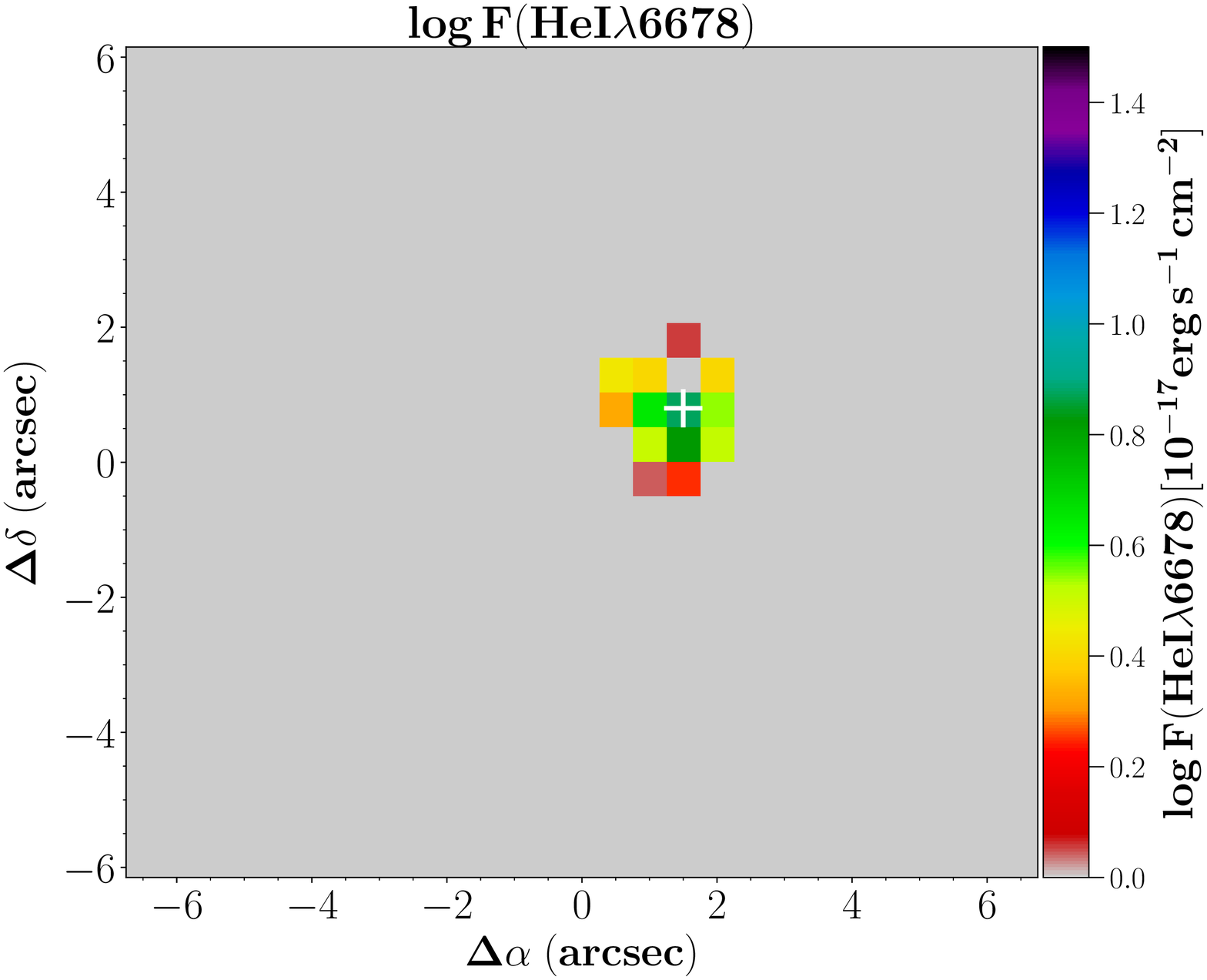}
\includegraphics[bb=4 4 793 648,width=0.43\textwidth,clip]{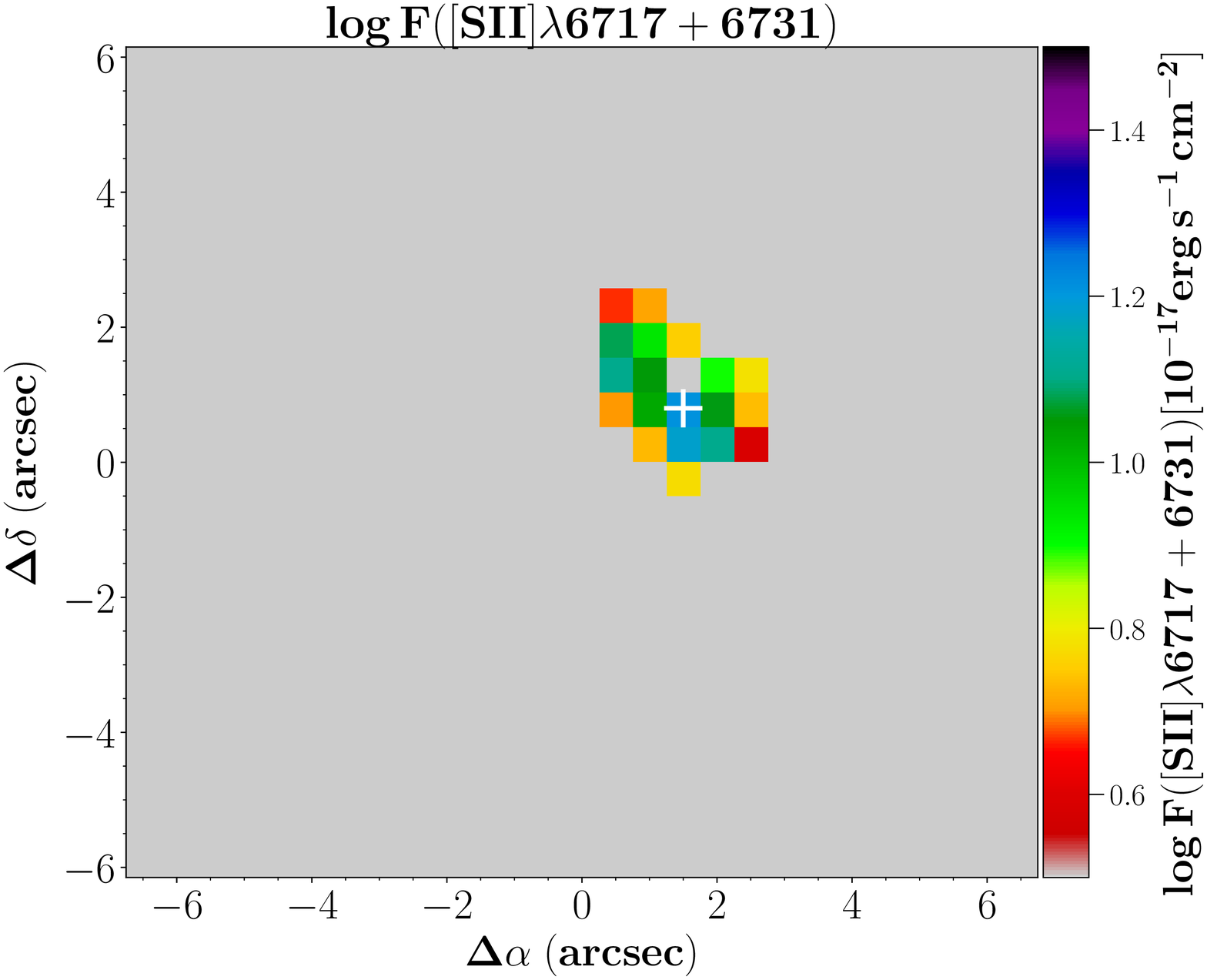}\\
\includegraphics[bb=4 4 793 648,width=0.43\textwidth,clip]{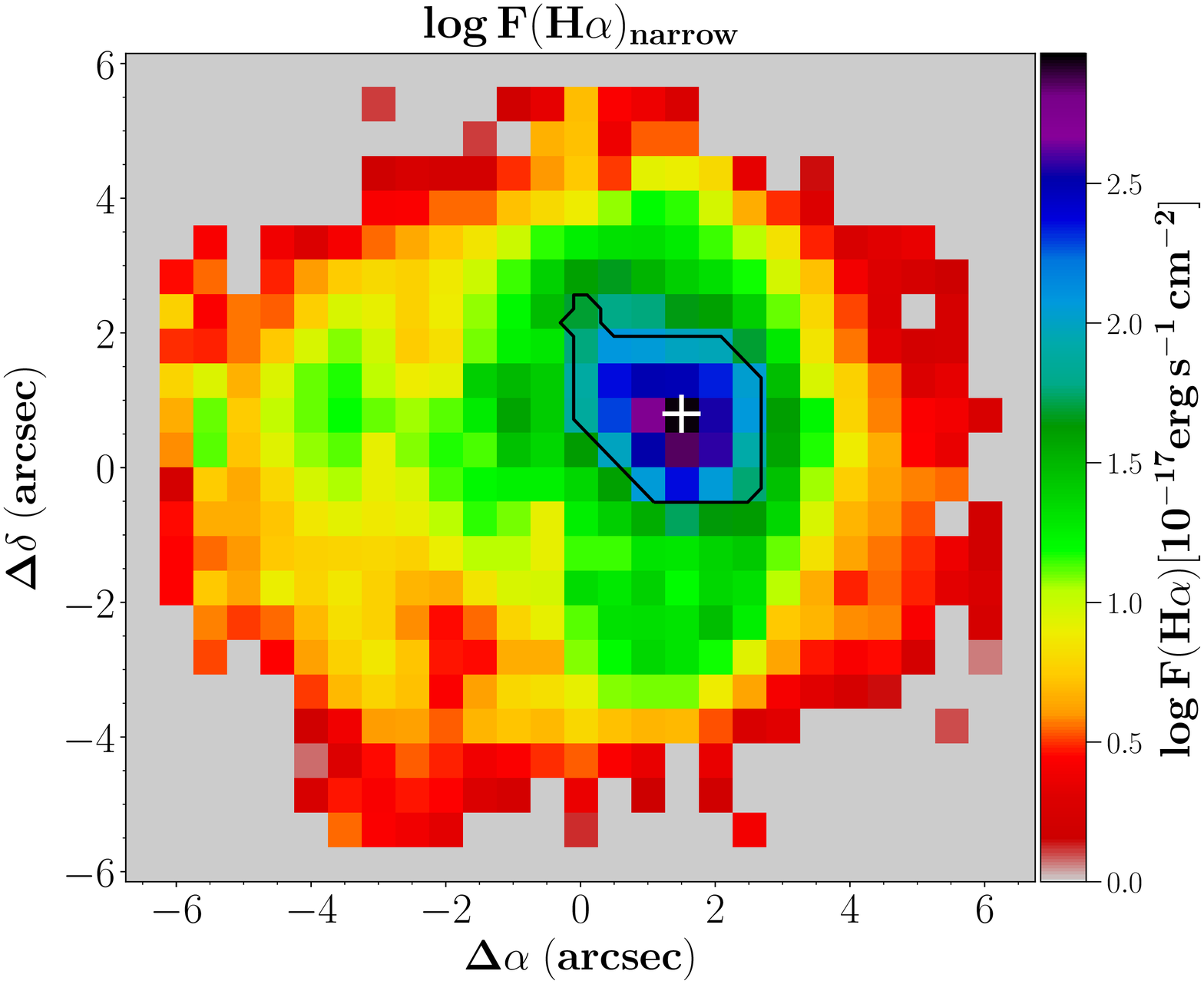}
\includegraphics[bb=4 4 793 648,width=0.43\textwidth,clip]{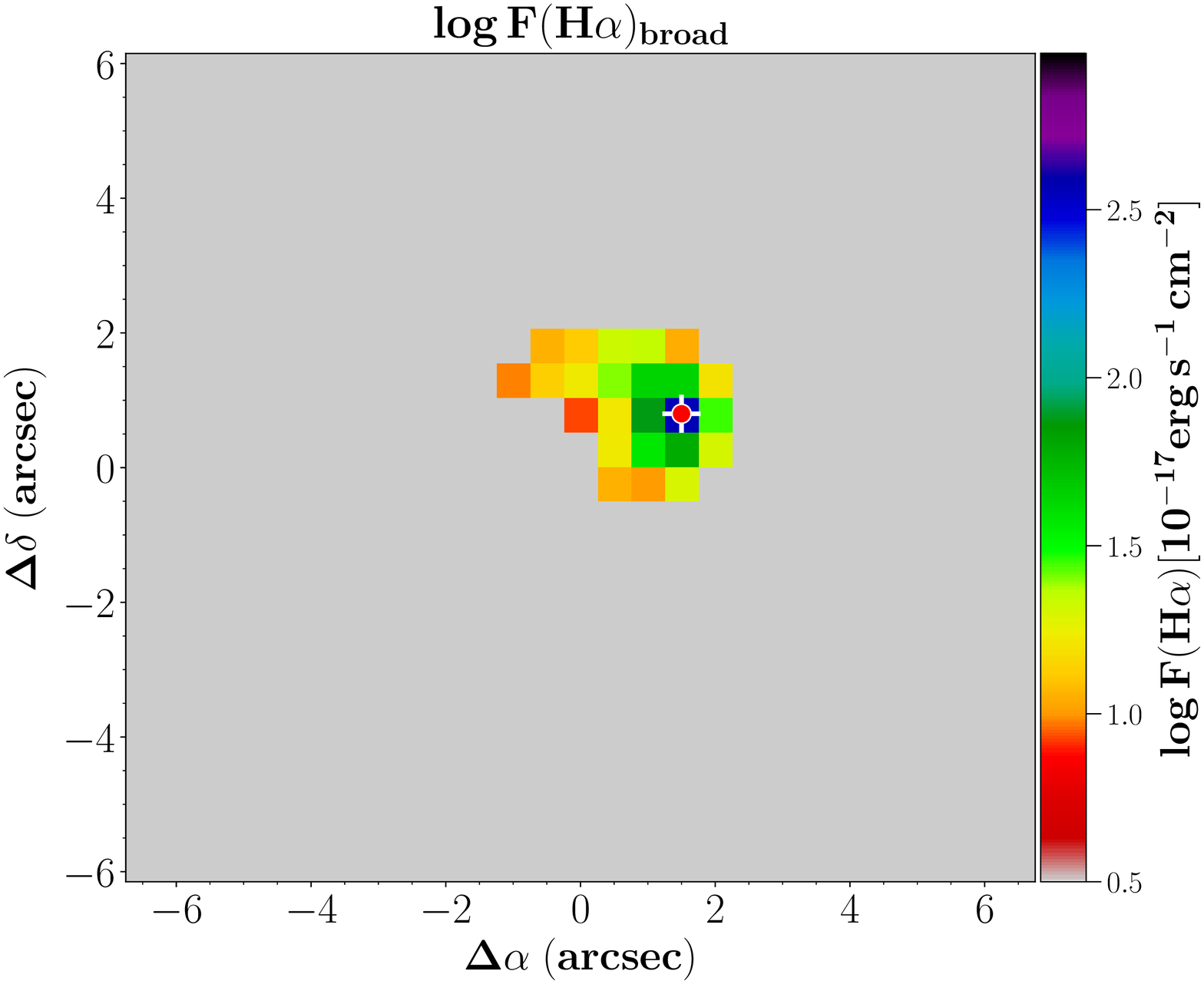}\\
\caption{(continue). The spaxel where we detect a P Cygni-like profile in the H$\alpha$ line is indicated with a red circle on the H$\alpha$ broad emission map; this spaxel corresponds to the H$\alpha$ emission peak. The black continuous line in the H$\alpha$ narrow emission map demarcates the brightest area of the galaxy enclosing only spaxels with H$\alpha$ S/N (per pixel) $>$ 100 (see Section~\ref{int} for details).}
\end{figure*}

\begin{figure*}
\center
\includegraphics[width=0.45\textwidth,clip]{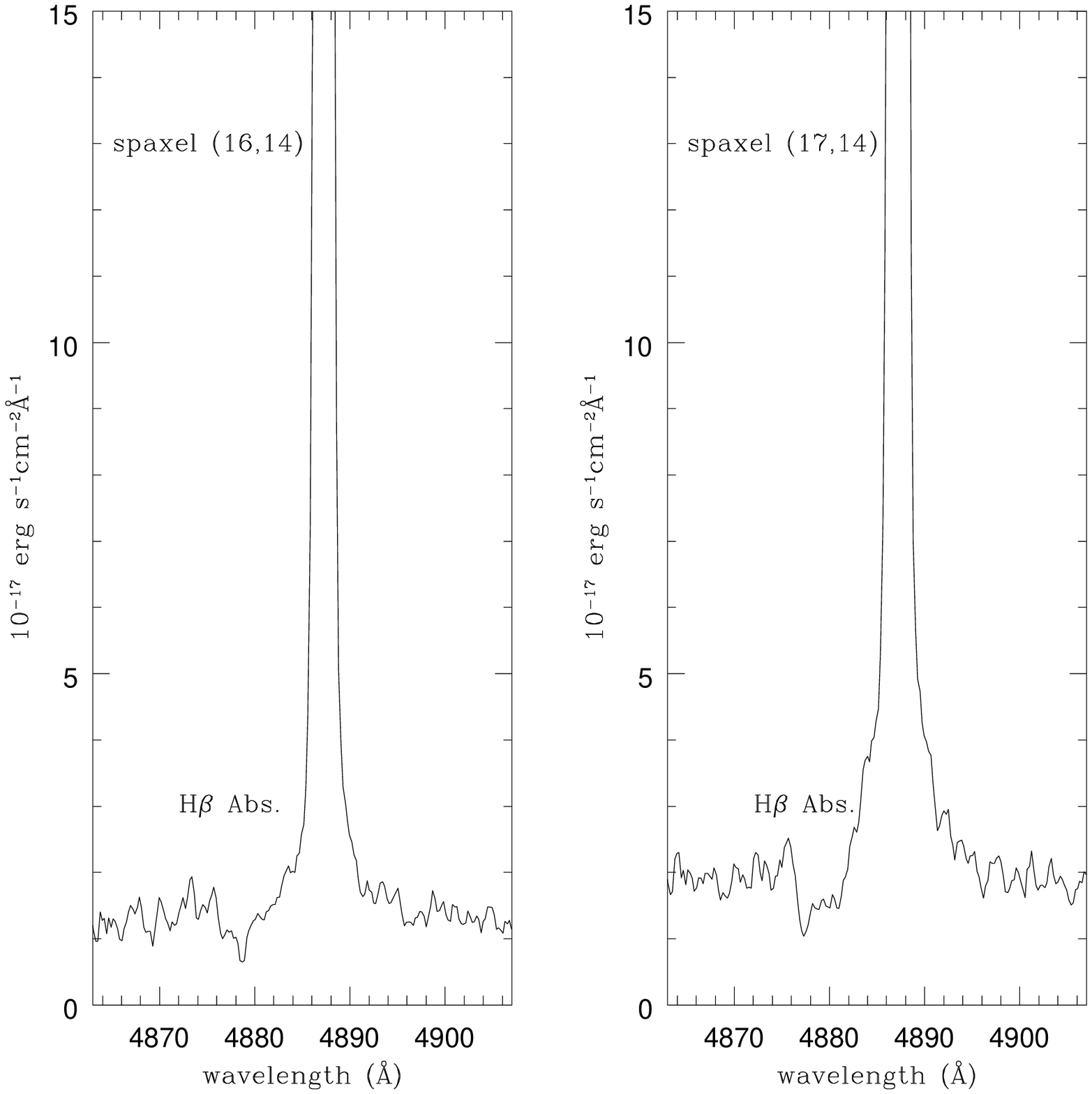}
\includegraphics[width=0.45\textwidth,clip]{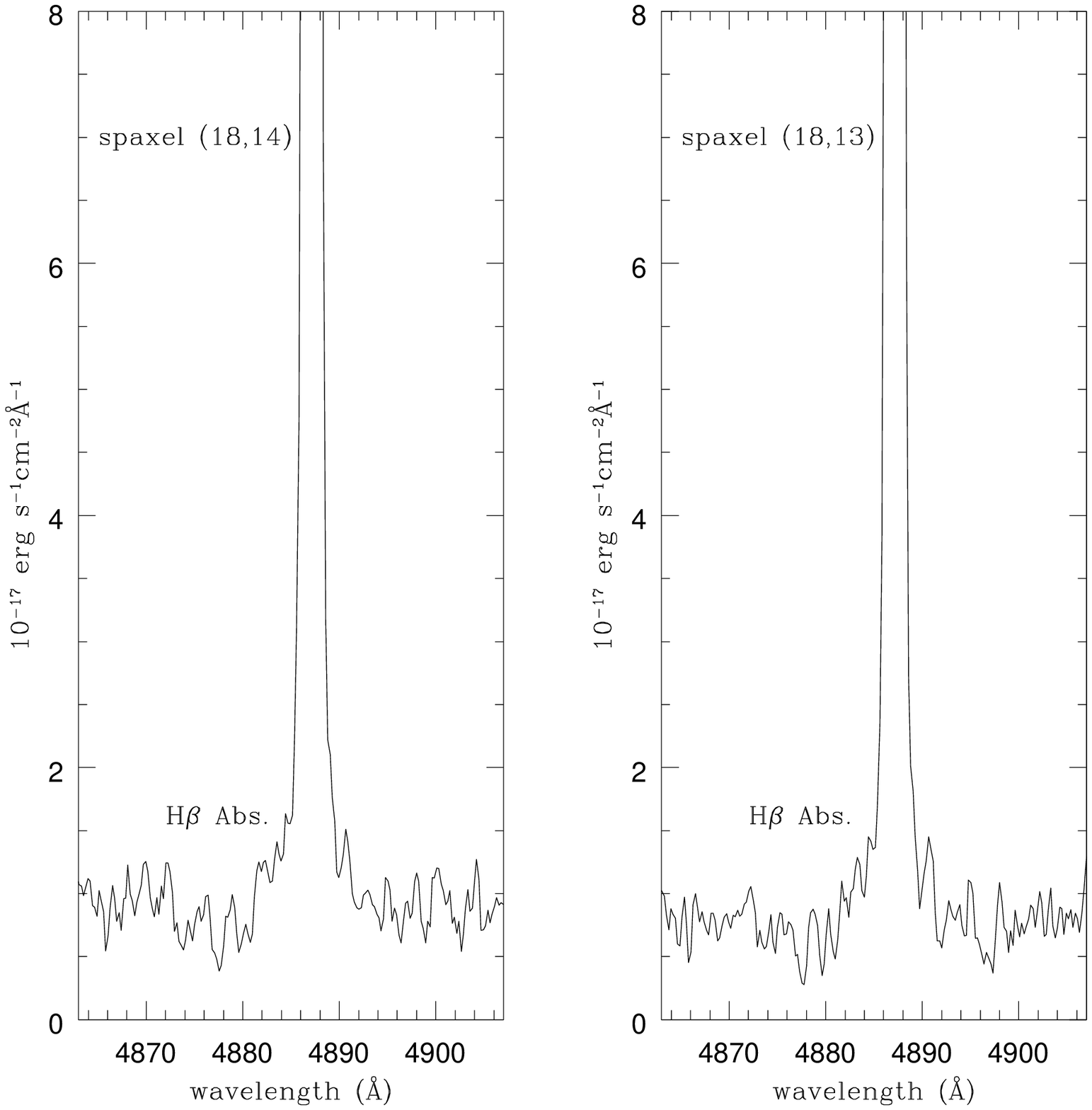}
\includegraphics[width=0.45\textwidth,clip]{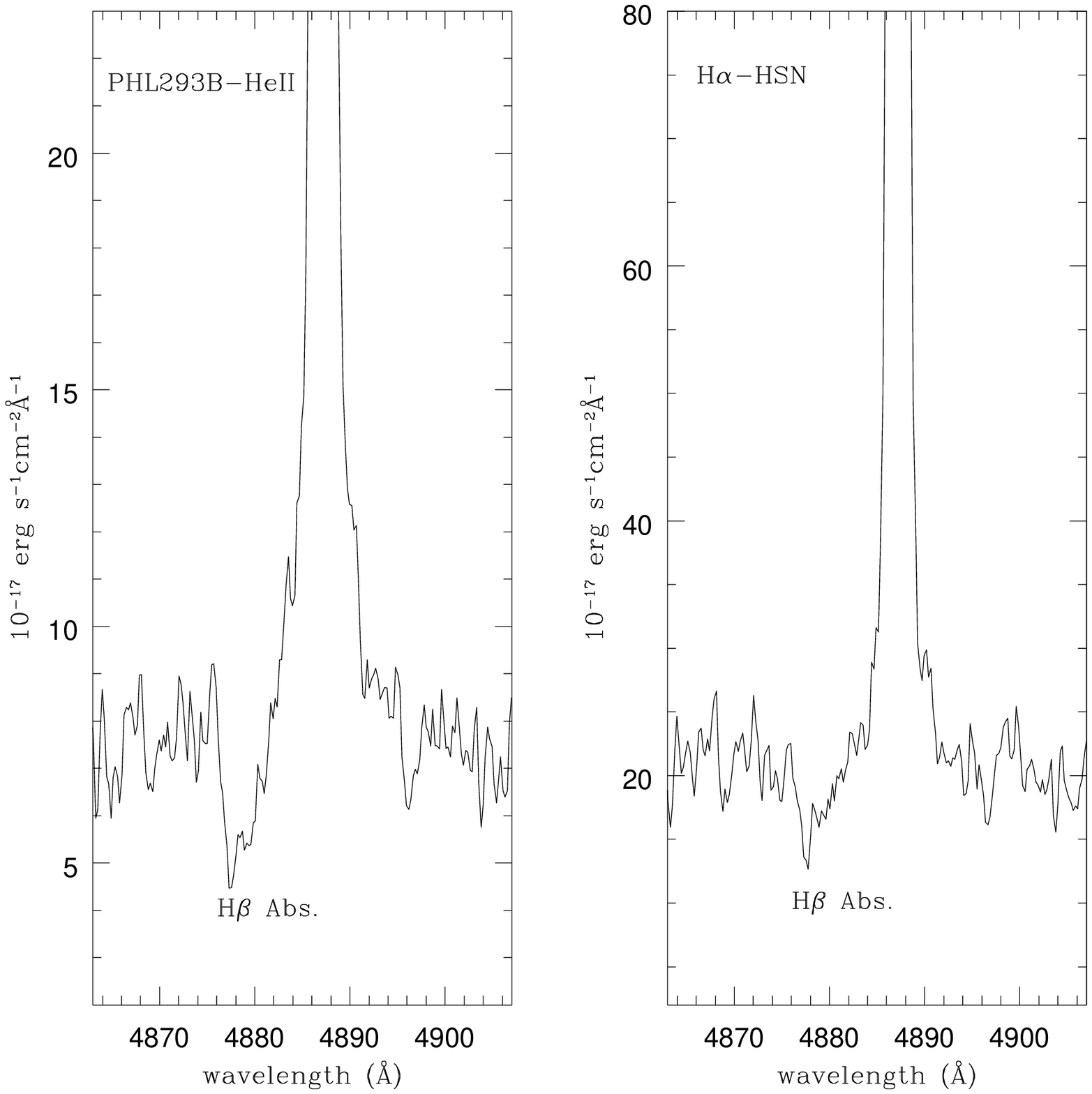}
\caption{Spectral regions around H$\beta$ where we detect P Cygni-like profiles. The {\it y-axis} shows the flux in units of 10$^{-17}$ erg s$^{-1}$ cm$^{-2}$ \AA$^{-1}$. {\it Top row}: the four individual spaxels marked with a black circle on the H$\beta$ broad emission map (see Fig.~\ref{line_maps}); for each spaxel we show in parenthesis the corresponding  cartesian coordinate (x,y) relative to the bottom left spaxel (0,0) in the maps. {\it Bottom row}: The spectra of the regions PHL~293B-HeII and H$\alpha$-HSN; the former is obtained by integrating the emission from all HeII$\lambda$4686-emitting spaxels, while the latter corresponds to the sum of all spaxels with H$\alpha$ S/N $>$ 100 (see Section~\ref{int} for details on the integrated spectra).} 
\label{pcyg_hb} 
\end{figure*}

\begin{figure}
\center
\includegraphics[bb=30 154 289 700,width=6cm,clip]{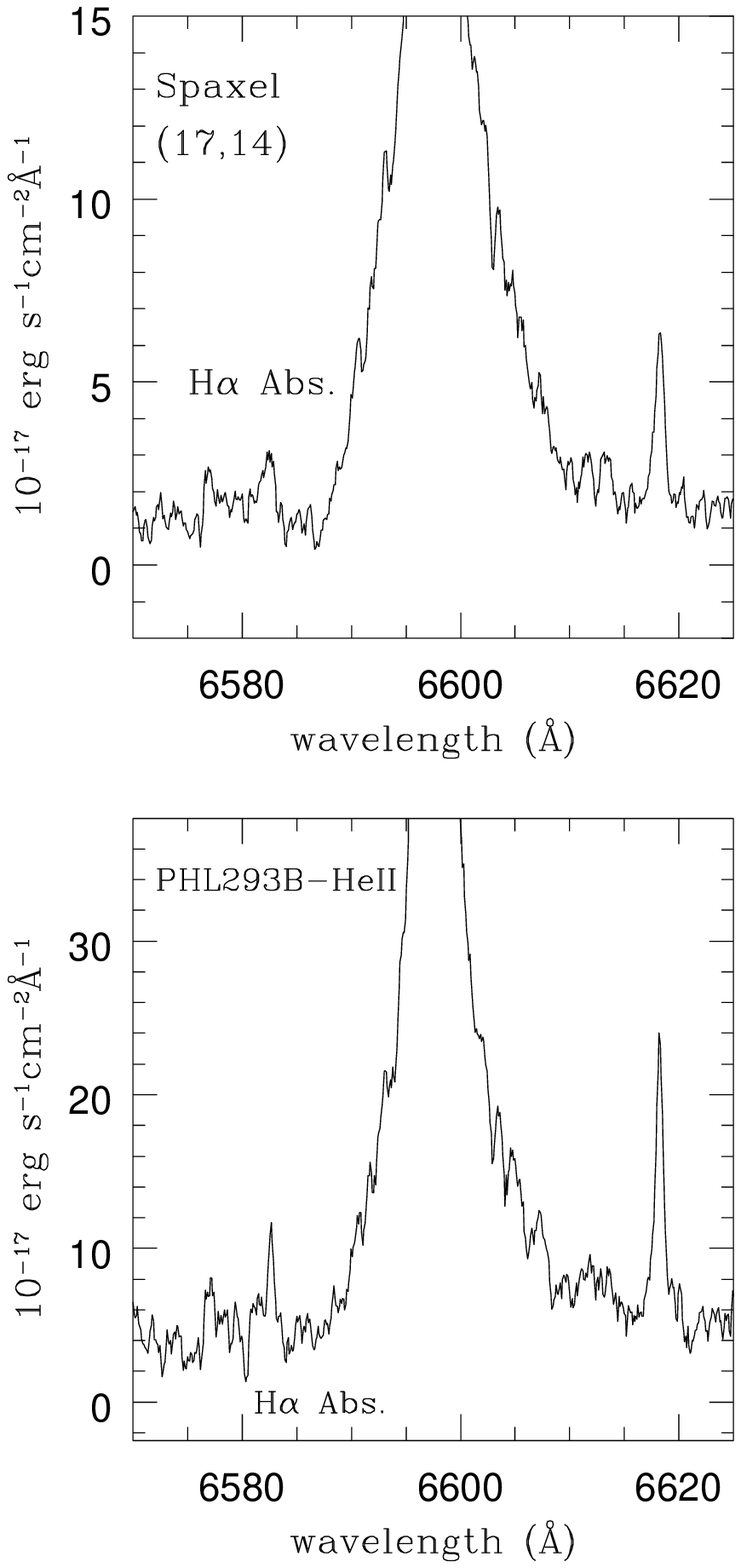}\\
\caption{Spectral regions around H$\alpha$ showing hints for P Cygni-like profiles. The {\it y-axis} shows the flux in units of 10$^{-17}$ erg s$^{-1}$ cm$^{-2}$ \AA$^{-1}$. {\it Top panel}: spaxel (17,14) which corresponds to the H$\alpha$ peak marked with a plus (+) sign on all maps; {\it Bottom panel}: Spectrum of the PHL~293B-HeII region as defined in the caption of Fig.~\ref{pcyg_hb} and in Section~\ref{int}.} 
\label{pcyg_ha} 
\end{figure}

\section{Spatially Resolved Properties of the ionized gas}
\label{2dgas}

\begin{figure} 
  \centering
\includegraphics[bb=25 25 596 464,width=0.43\textwidth,clip]{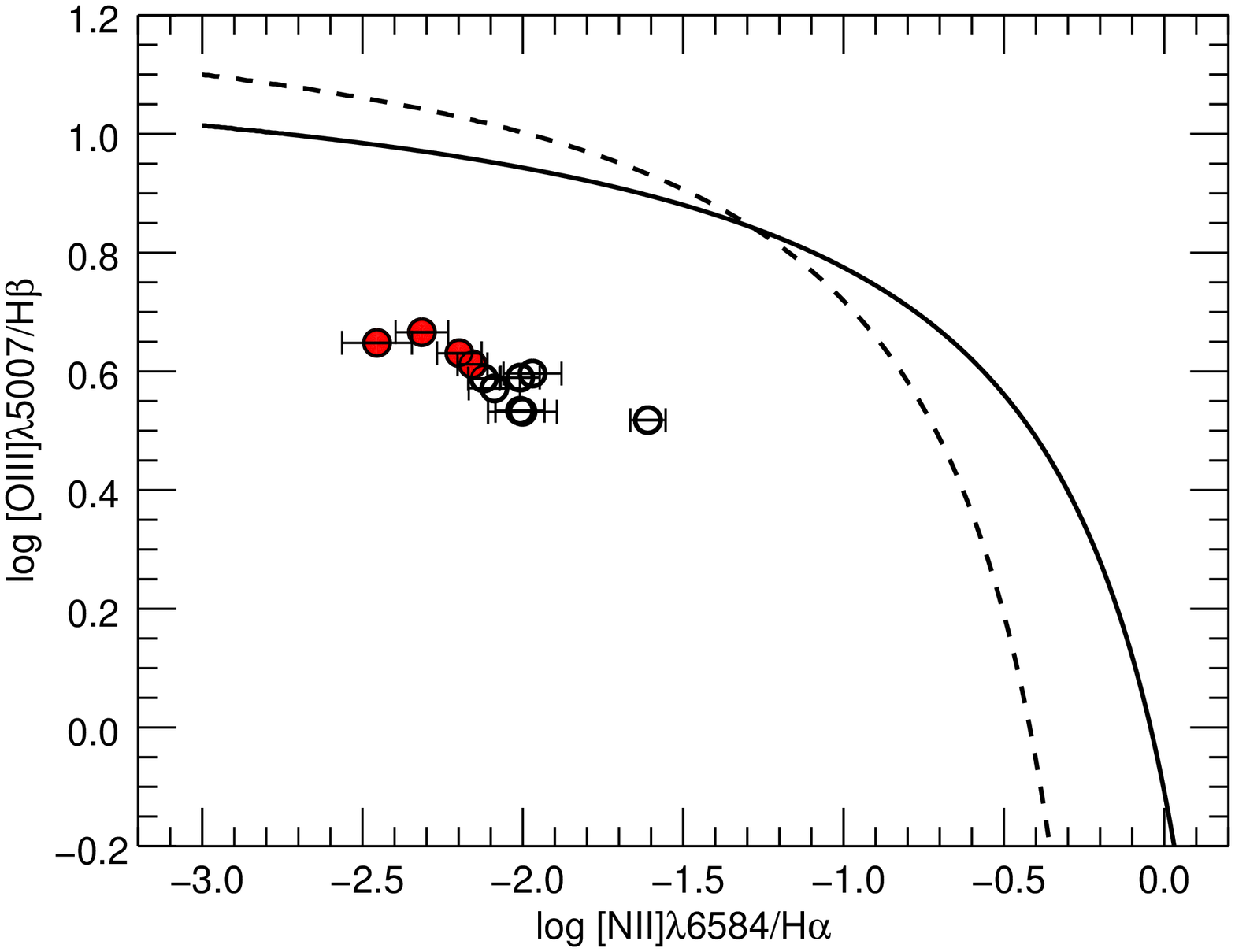} 
\includegraphics[bb=25 25 596 464,width=0.43\textwidth,clip]{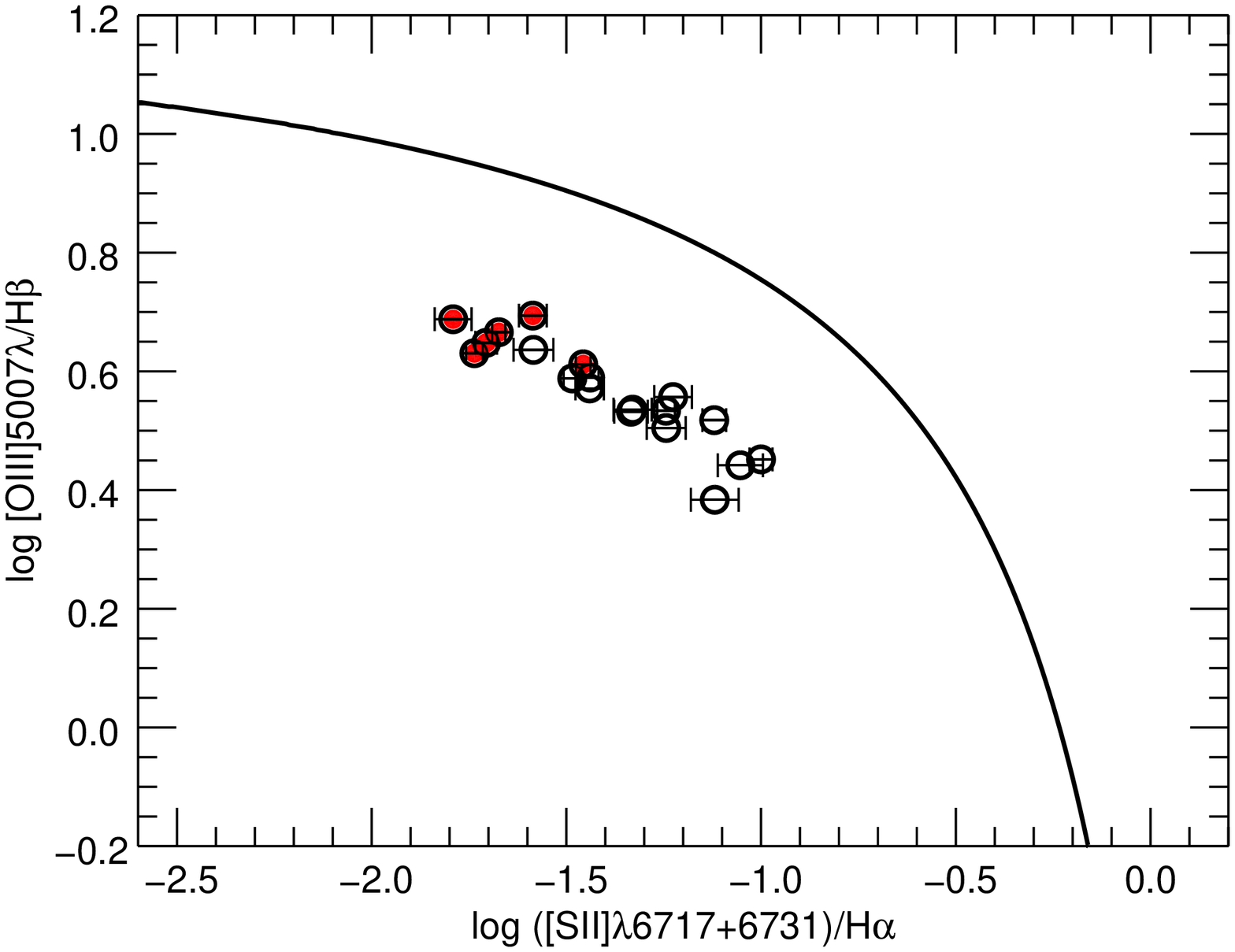} 
\caption{BPT diagnostic diagrams for PHL~293B. Open circles
mark the individual spaxels from the data cube; red circles show
the individual HeII-emitting spaxels.The solid line (in the two panels) indicates the theoretical demarcation limit from Kewley et al. (2001) that separates objects where the gas ionization is  mainly due to hot massive stars (below and to the left of the curve) from those where other ionizing mechanism is required. The dashed line in the [N{\sc ii}]$\lambda$6584/H$\alpha$ plot (top panel) depicts the boundary between SF systems and AGNs from Kauffmann et al. (2003). Error bars in the {\it y-axis} are the same size or smaller than the symbols, and are not plotted for the sake of clarity.}
\label{bpt} 
\end{figure} 

\subsection{Ionization Structure}

Baldwin-Phillips-Terlevich (BPT) diagrams \citep[][]{bal81} are a
powerful tool, widely used to separate star-forming galaxies and AGN. The spatially-resolved BPT diagrams for PHL~293B are shown in Fig.~\ref{bpt}:
[O{\sc iii}]$\lambda$5007/H$\beta$ vs. [N{\sc
ii}]$\lambda$6584/H$\alpha$, [S{\sc
ii}]$\lambda\lambda$6717,6731/H$\alpha$. These line ratios are not corrected for extinction, but reddening effects must be minor since these ratios involve lines which are close in wavelength. Each circle plotted in Fig.~\ref{bpt} corresponds to a line ratio obtained from a single spaxel, where the red circles show the HeII$\lambda$4686-emitting spaxels. Based on stellar population synthesis and photoionization models, \cite{kew01} proposed a 
theoretical demarcation curve that isolates galaxies with line ratios which are due to excitation by massive stars within HII
regions from those where other ionizing source is needed. An
empirical curve that differentiates between AGNs and HII-like systems was later derived by \cite{K03}; both demarcation lines are plotted in Fig.~\ref{bpt}.  For all positions in PHL~293B our emission-line
ratios fall in the general locus of SF objects, i.e., below and to the left of the separation lines in the two BPT diagrams. This suggests that
photoionization from hot massive stars appears to be the dominant excitation
mechanism within PHL~293B.

\begin{figure*}
\center
\includegraphics[bb=4 4 793 648,width=0.45\textwidth,clip]{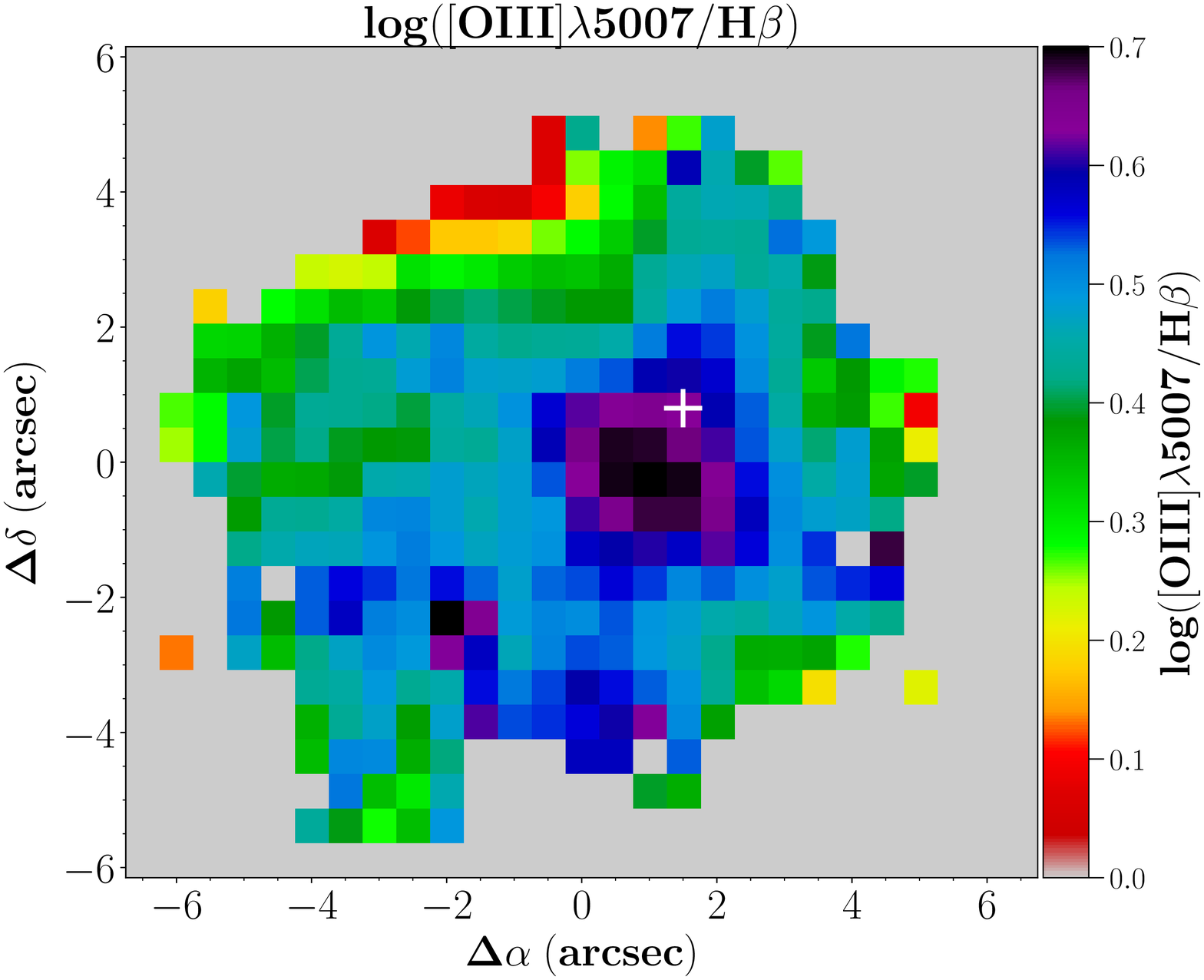}
\includegraphics[bb=4 4 793 648,width=0.45\textwidth,clip]{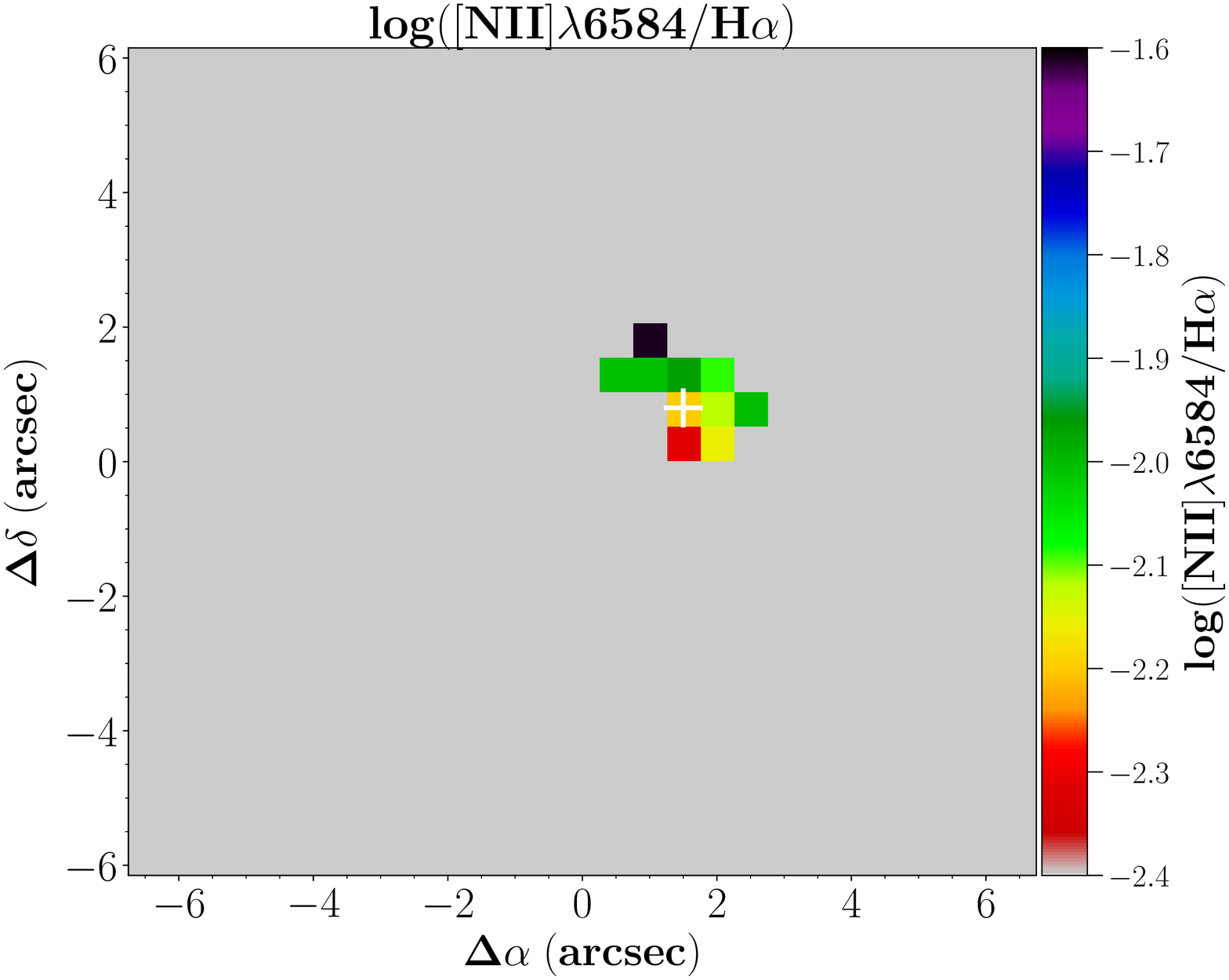}\\
\includegraphics[bb=4 4 793 648,width=0.45\textwidth,clip]{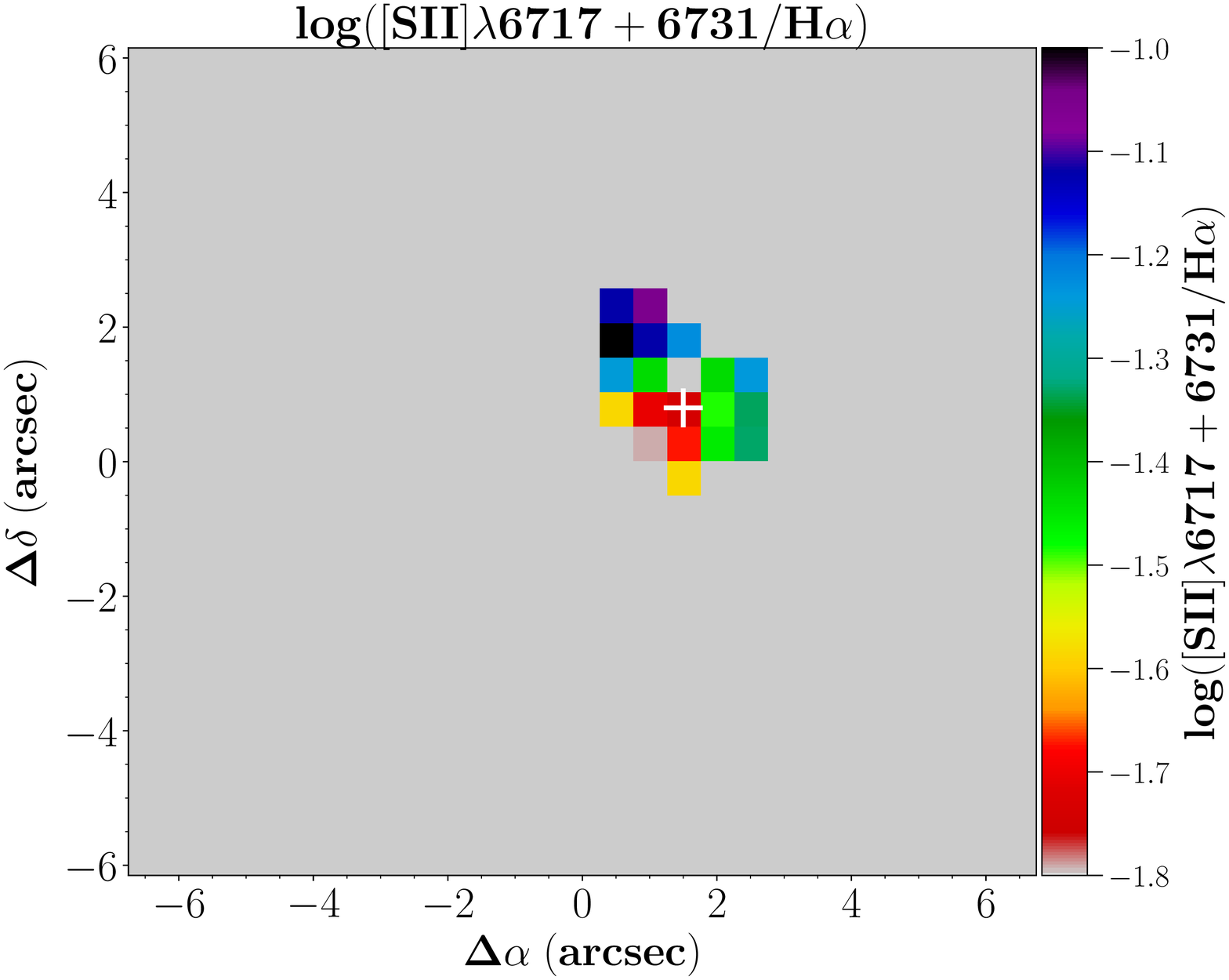}
\caption{Maps of line ratios in logarithmic scale. East is left and North is up. The plus (+) sign is as indicated in} Fig.~\ref{line_maps}.
\label{line_ratio_maps} 
\end{figure*}

The spatial distribution for the BPT line ratios are displayed in
Fig~\ref{line_ratio_maps}. While highest and lowest values of [O{\sc iii}]$\lambda$5007/H$\beta$
are found at the most inner and external zones of PHL~293B
respectively, a reverse trend is observed in the [N{\sc
ii}]$\lambda$6584/H$\alpha$ and [S{\sc
ii}]$\lambda\lambda$6717,6731/H$\alpha$ maps, indicating the presence
of higher excited gas inward. The [O{\sc iii}]$\lambda$5007/H$\beta$
map clearly shows larger values spatially coincident with the southern
HII region (i.e., the bright blue knot in Fig.~\ref{megara_ifu}) which also
comprises the HeII zone (see also the HeII$\lambda$4686 map in
Fig.~\ref{line_maps}). Additionally, in Fig.~\ref{bpt} we see that the
HeII-emitting spaxels tend to have higher [O{\sc
iii}]$\lambda$5007/H$\beta$ ratio in comparison to the other
spaxels. Shall we note that despite this correlation, the hard HeII-ionizing radiation (E $\gtrsim$ 4
Ryd) does not have to be necessarily the main responsible for the
brighter [O{\sc iii}] emission since the the [O{\sc iii}] lines can be
excited by softer energies (E $\gtrsim$ 2.5 Ryd) \citep[see e.g.,][]{TI05}.

\begin{figure*}
\center
\includegraphics[bb=4 4 793 648,width=0.45\textwidth,clip]{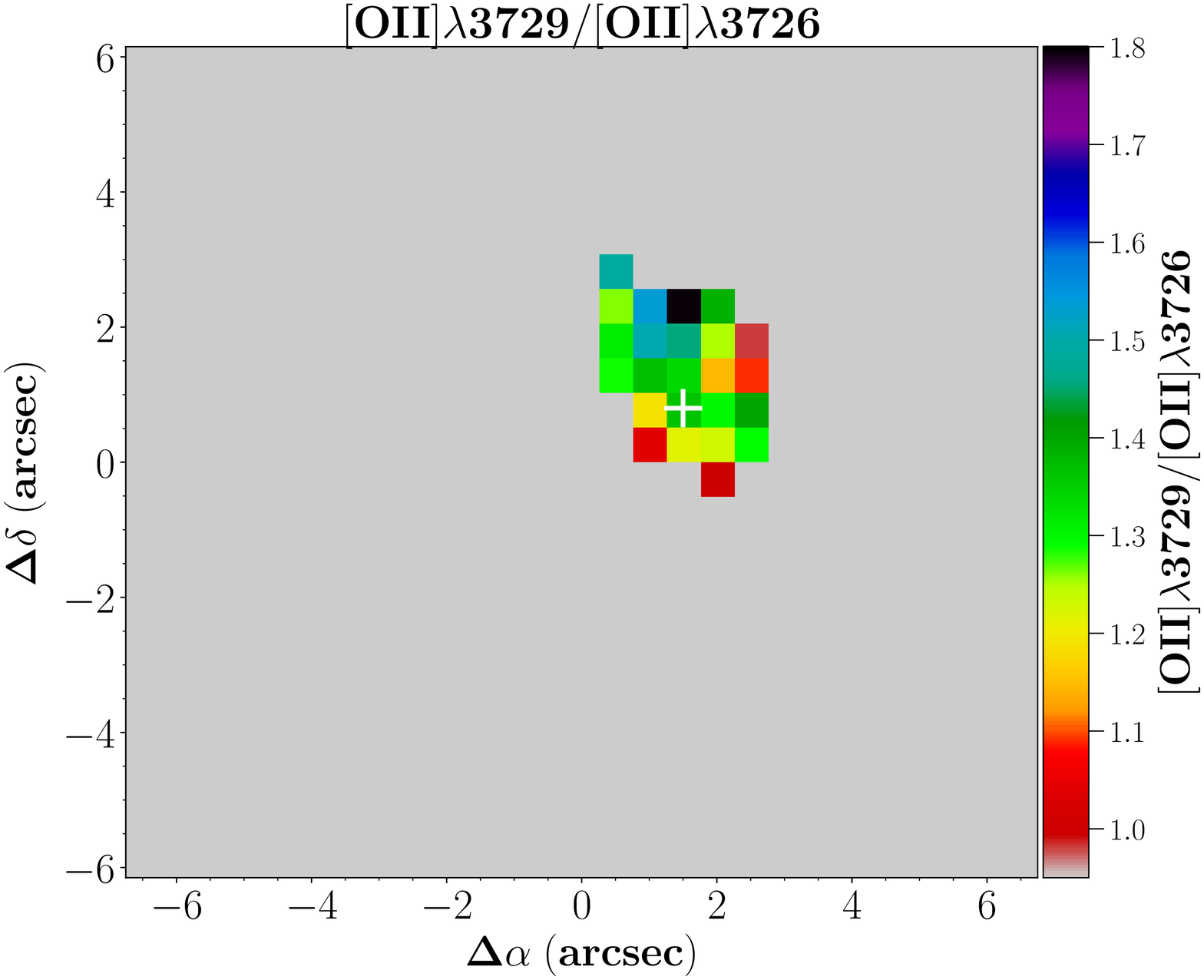}
\includegraphics[bb=4 4 793 648,width=0.45\textwidth,clip]{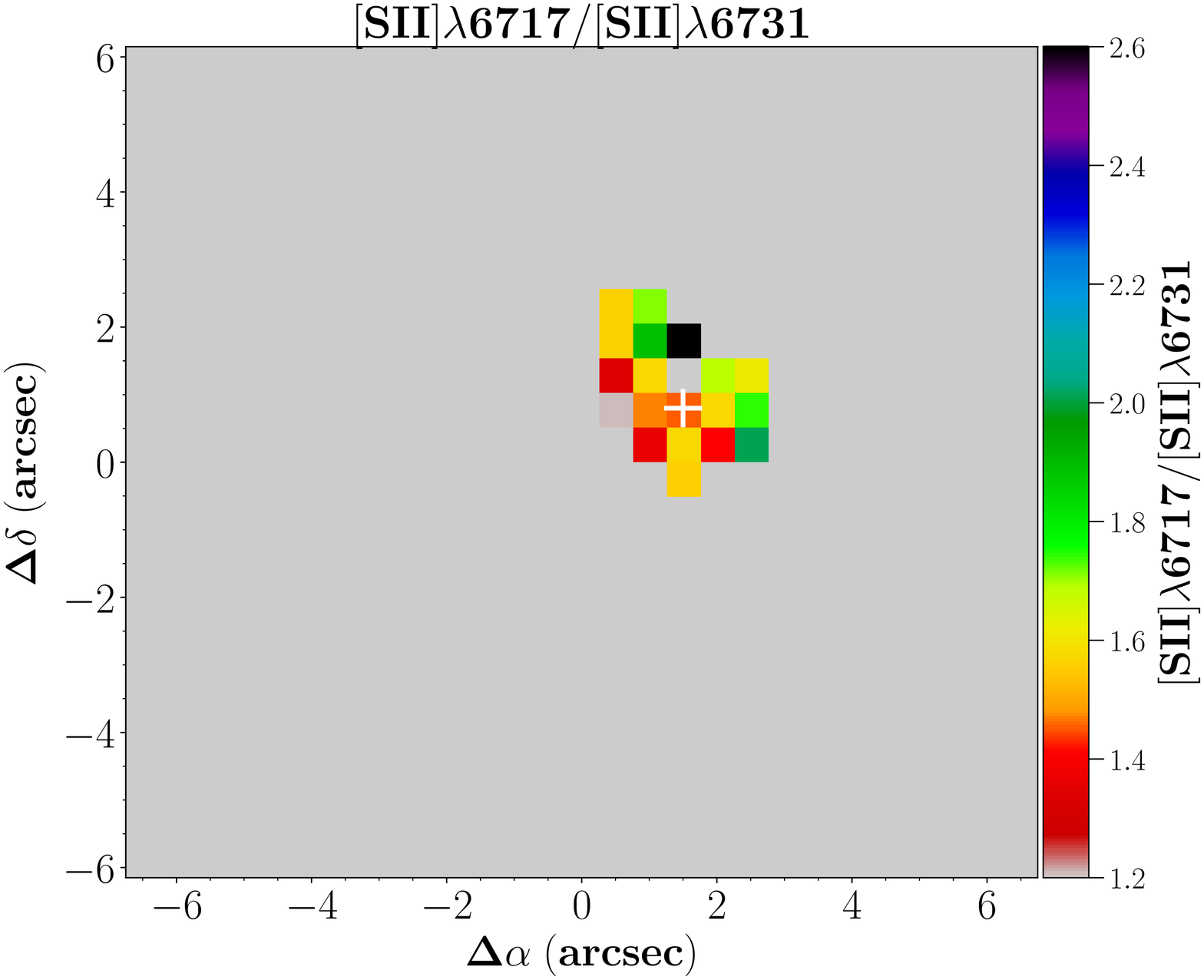}\\
\caption{Indicators of nebula electron density. Maps for {\mbox [O{\sc ii}]$\lambda$3729/[O{\sc ii}]$\lambda$3726} and {\mbox [S{\sc ii}]$\lambda$6716/[S{\sc ii}]$\lambda$6731}. East is left and North is up. The plus (+) sign is as indicated in} Fig.~\ref{line_maps}.
\label{density} 
\end{figure*}

\begin{figure*}
\center
\includegraphics[bb=4 4 793 648,width=0.45\textwidth,clip]{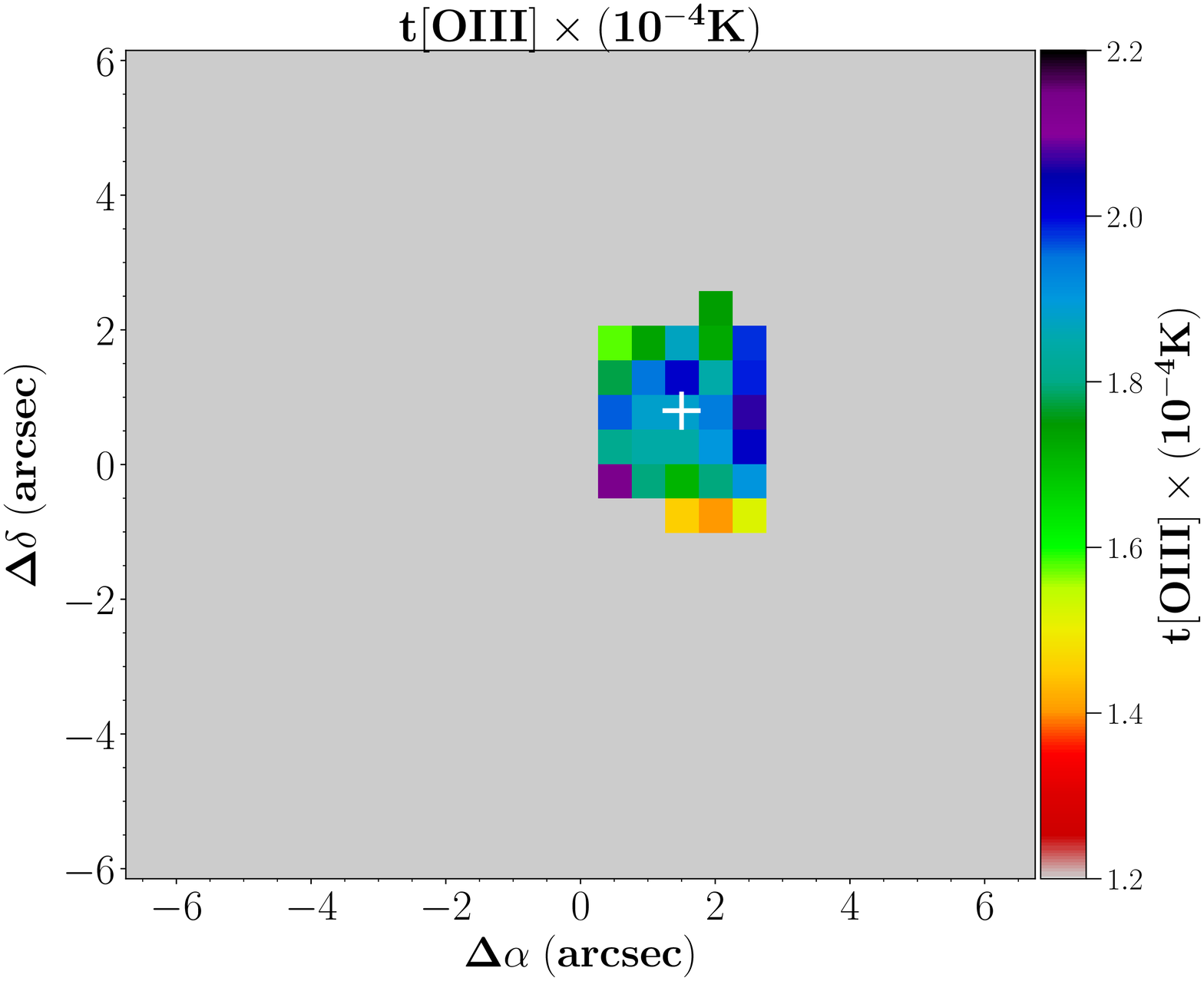}
\includegraphics[bb=4 4 793 648,width=0.45\textwidth,clip]{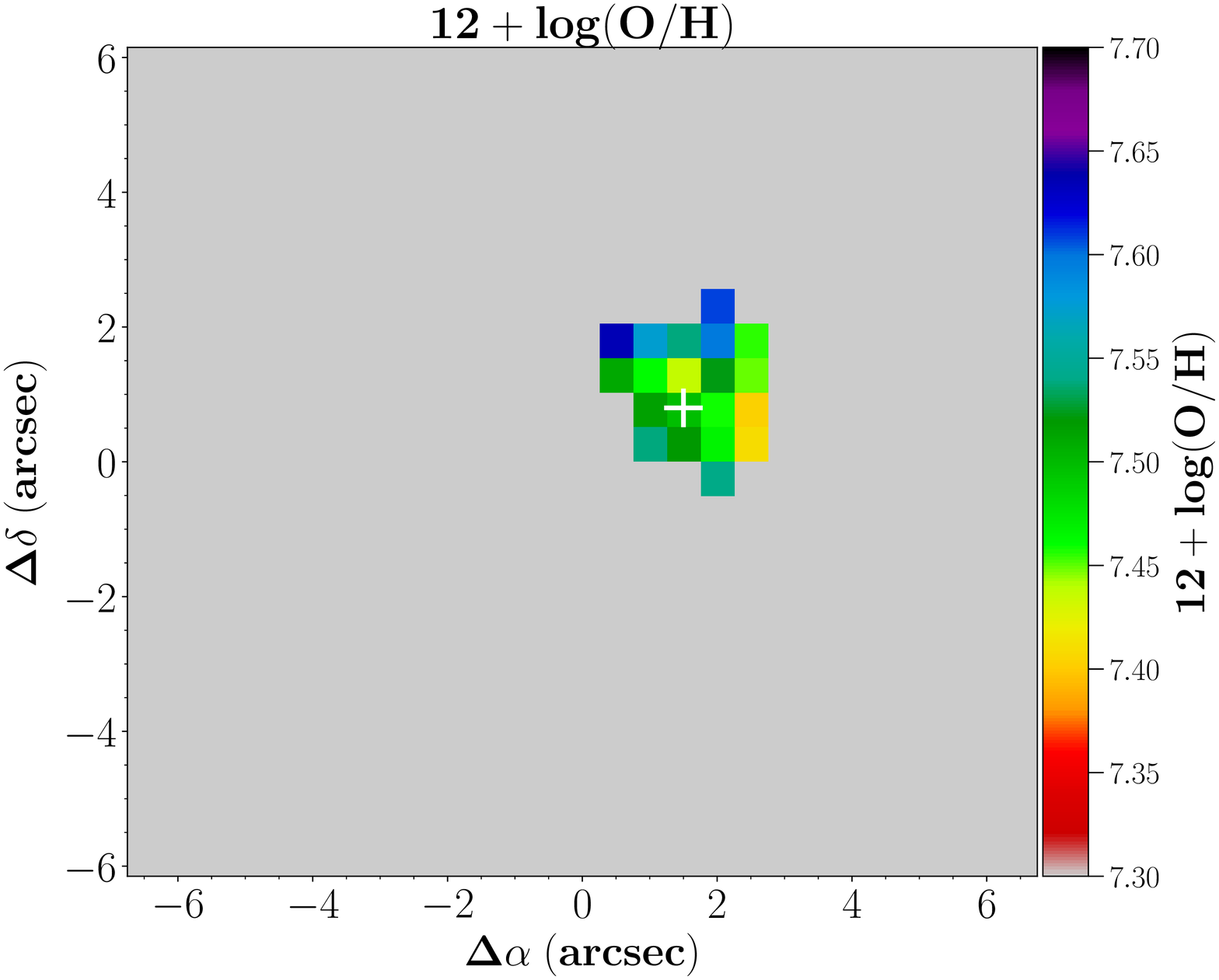}\\
\includegraphics[bb=19 280 605 465,width=0.75\textwidth,clip]{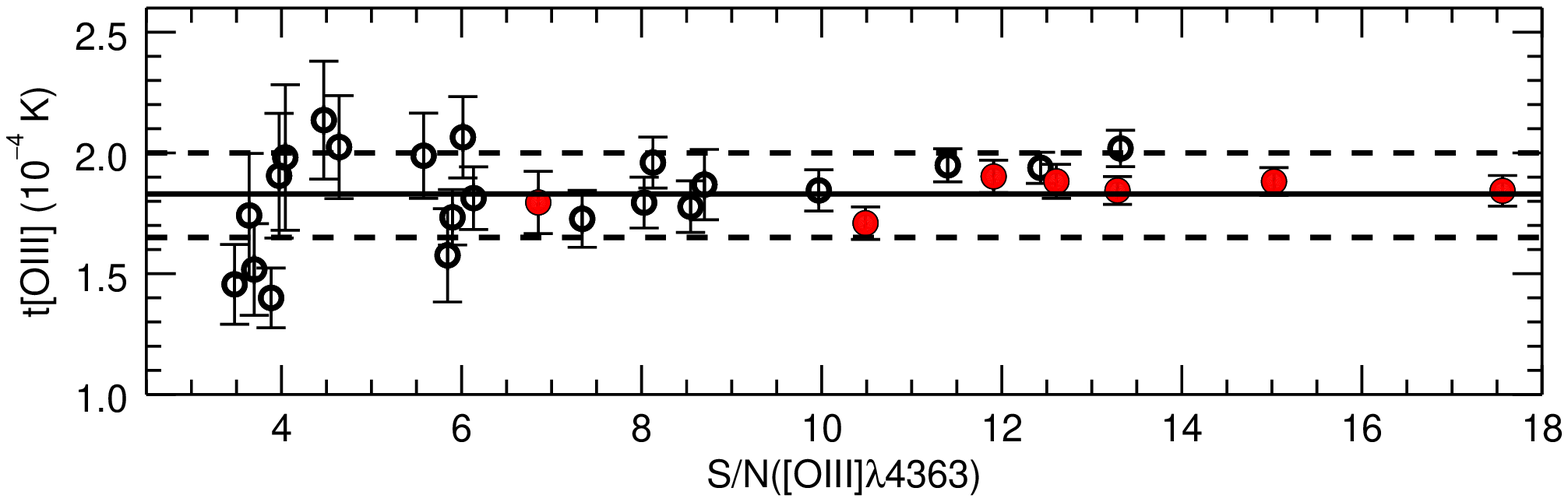}
\caption{{\it Top-left panel}: Map of the [O{\sc iii}]$\lambda$4363-derived T$_{e}$[O{\sc iii}]. {\it Top-right panel}: Map of oxygen abundance derived only for spaxels with T$_{e}$[O{\sc iii}] measurements available. The plus (+) sign is as indicated in Fig.~\ref{line_maps}. {\it Bottom panel}: T$_{e}$[O{\sc iii}] versus the S/N of the [O{\sc iii}]$\lambda$4363 line flux. Open circles represent individual spaxels; red circles indicate the seven HeII$\lambda$4686-emitting spaxels; the solid horizontal line marks the mean value for $T_{\rm e}$[O{\sc iii}] $\sim$ 1.83 $\times$ 10$^{4}$ K, while dotted lines represent $\pm$ 1$\sigma$.} 
\label{te} 
\end{figure*}

\subsection{Nebular physical-chemical properties on a spaxel-by-spaxel basis}

We have used the expressions from \citet{EPM17} to compute the
physical properties and ionic abundances of the PHL~293B ionized
gaseous nebulae. These expressions are derived from the PyNEB tool \citep{L15}.

In Fig.~\ref{density} we show the maps for the {\mbox [O{\sc
ii}]$\lambda$3729/$\lambda$3726} and {\mbox [S{\sc
ii}]$\lambda$6717/$\lambda$6731} line ratios which are good indicators
of the average electron density ($n_{\rm e}$) in a
nebula \citep{OF06}. For most of the spectra, the observed [O{\sc ii}]
([S{\sc ii}]) line ratios correspond to $n_{\rm e}$ values
$\lesssim$ 300 cm$^{-3}$ ($\lesssim$ 100 cm$^{-3}$), indicating a
relatively low-density ionized gas in the central parts of PHL~293B.

For the [O{\sc iii}]$\lambda$4363-emitting spaxels, we have computed
the electron temperature $T_{\rm e}$[O{\sc iii}] values from the
reddening corrected [O{\sc iii}]$\lambda$4363/[O{\sc
iii}]$\lambda$5007 line ratio. We have measured the weak [O{\sc
iii}]$\lambda$4363 line above 3$\sigma$ for 29 spaxels which extend to
an area of around 9.6 arcsec$^{2}$ equivalent to 0.12 kpc$^{2}$ (see
Fig.~\ref{line_maps}). The top-left panel of Fig.~\ref{te} presents
the map of the $T_{\rm e}$[O{\sc iii}] which reveales values going
from $\gtrsim$ 14,000 K to near 20,000 K, with a good fraction of the
points clustering around the average $T_{\rm e}$ value of $\sim$
18,300 K. The relation between $T_{\rm e}$[O{\sc iii}] and the S/N
measured for the [O{\sc iii}]$\lambda$4363 line is plotted in the
bottom panel of Fig.~\ref{te} where no systematic effects are
observed. This is an evidence that the largest values of $T_{\rm
e}$[O{\sc iii}] that we derive are not an effect of overestimated
[O{\sc iii}]$\lambda$4363 flux measurements. We have used
the \mbox{$T_{\rm e}$[O{\sc ii}]-$T_{\rm e}$[O{\sc iii}]} empirical
relantionship from \cite{PI06} to determine the $T_{\rm e}$[O{\sc ii}]
values since no low-excitation auroral line (e.g., [N{\sc
ii}]$\lambda$5755) has been detected in any spaxel.

The O$^{+}$/H$^{+}$ and O$^{2+}$/H$^{+}$ ionic abundance ratios, were
computed from the [O{\sc ii}]$\lambda$3726,29 and \mbox{[O{\sc
iii}]$\lambda$5007} lines, respectively using the corresponding
electron temperatures. A tiny fraction of the unseen O$^{3+}$ ion is
expected to be present in high-ionizing SF regions as the ones showing
HeII emission. Based on the photoionization models from \citet{I06},
the O$^{3+}$/O ratio is $>$ 1$\%$ only in the highest-excitation HII
regions whose O$^{+}$/(O$^{+}$ + O$^{2+}$) is lower than 10$\%$. We
have checked that for all [OIII]$\lambda$4363-emitting spaxels
(including non-HeII$\lambda$4686 and HeII$\lambda$4686 emitting
spaxels), O$^{+}$/(O$^{+}$ + O$^{2+}$) is $\gtrsim$ 10$\%$; therefore
the total O/H is assumed to be O$^{+}$/H$^{+}$ + O$^{2+}$/H$^{+}$. The
spatial distribution of the derived 12+log(O/H) is displayed in the
top-right panel of Fig.~\ref{te} with most of the spaxels (80$\%$)
showing oxygen abundance in the range \mbox{of $\approx
7.5$--$7.6$}. Our results, thus, indicate that the warm gas-phase O/H
in PHL~293B stands largely constant beyond hundreds of parsecs. This
agrees with the absence of significant abundance gradient commonly
observed in the ionized gas of HII
galaxies \citep[e.g.,][]{KS96,I99,T99,I01,I04,I06b,P06,K08,K13,K16,EPM09,EPM11}.

\section{Integrated spectra across the MEGARA FOV of PHL~293B}\label{int}

Based on our IFU data we integrated individual spectra of selected
galaxy regions. We created for the first time the integrated spectrum
of PHL~293B by adding the flux in all the spaxels with H$\alpha$ S/N
(per spaxel) $>$ 3; this matches an area of $\sim$ 194 arcsec$^{2}$
($\sim$ 2.4 kpc$^{2}$) enclosing basically all the nebular emission
across our FOV. In addition, by summing the emission from the spaxels
with H$\alpha$ S/N $>$ 100 ($\sim$ 11.5 arcsec$^{2}$), we simulate the
spectrum of the brightest region of the galaxy (hereafter
H$\alpha$-HSN region\footnote{HSN from High Signal to Noise ratio.}),
and whose boundary is shown overplotted on the map of H$\alpha$ (see
Fig.~\ref{line_maps}). Finally, we obtained the spectrum of the region
that we name PHL~293B-HeII. To do so we have integrated all
HeII-emitting spaxels which covers $\sim$ 3.5 arcsec$^{2}$ (see the
HeII map in Fig.~\ref{line_maps}).

\begin{figure*}
\center
\includegraphics[width=0.49\textwidth,clip]{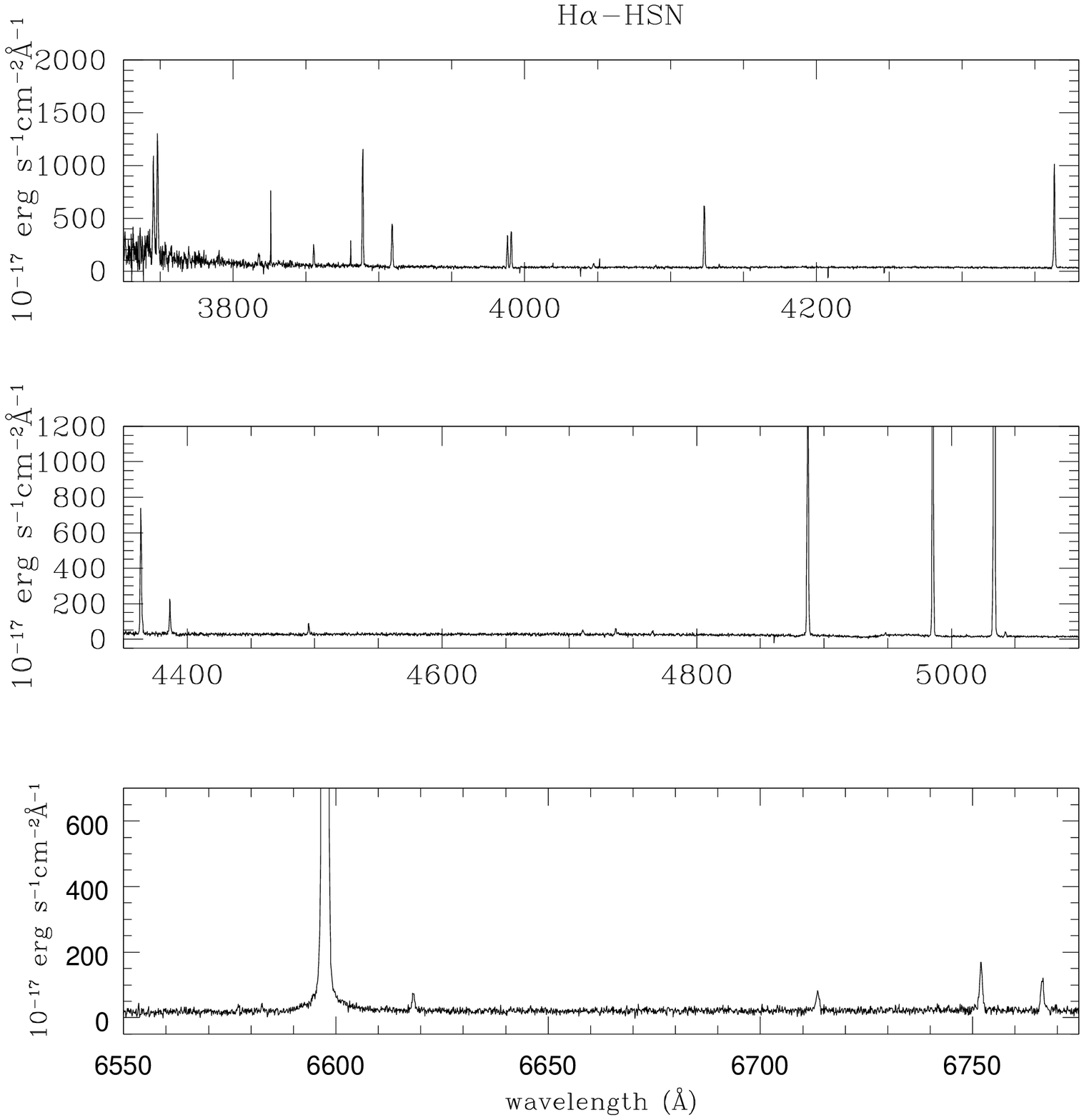}
\includegraphics[width=0.49\textwidth,clip]{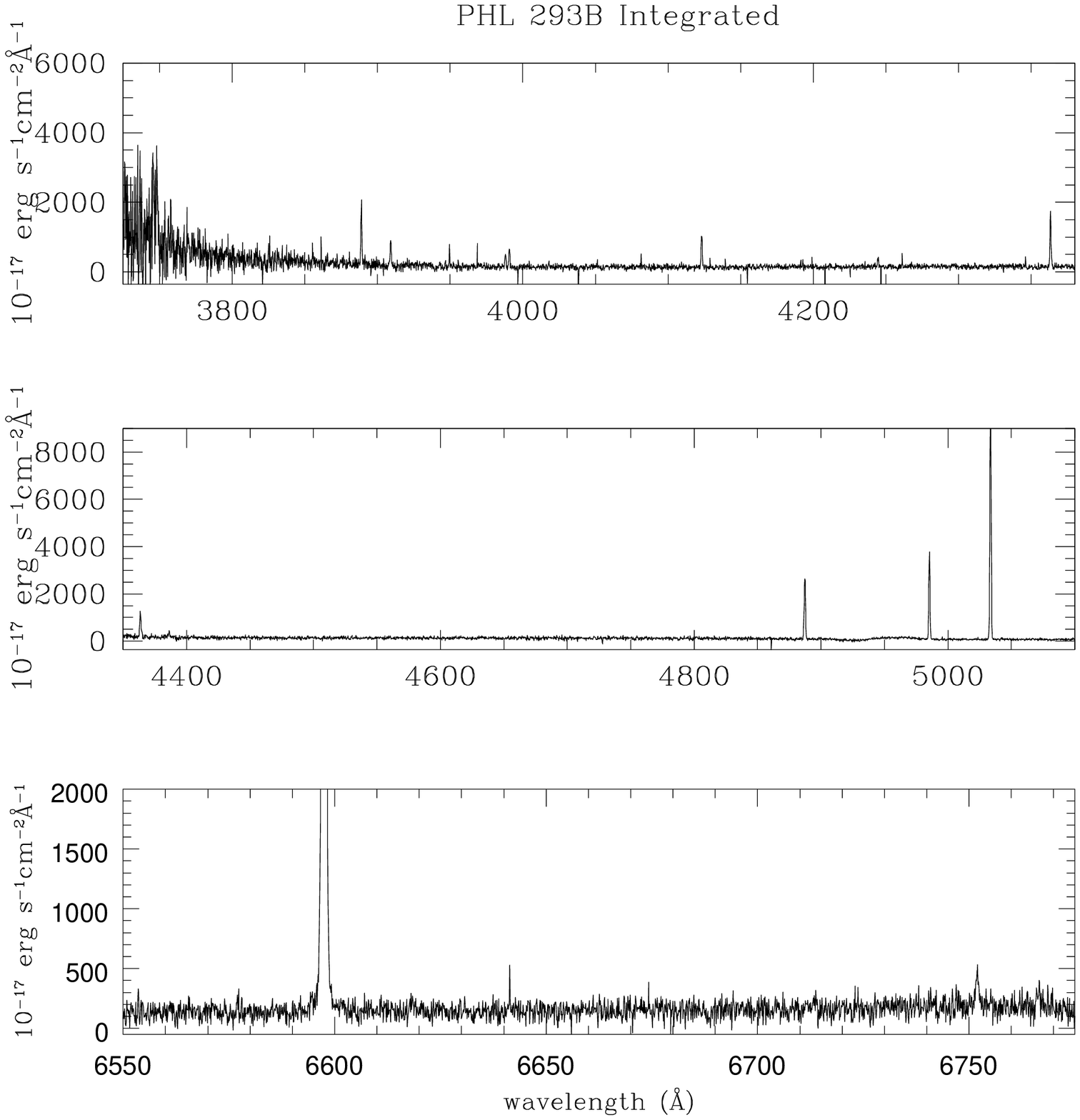}\\
\includegraphics[width=0.49\textwidth,clip]{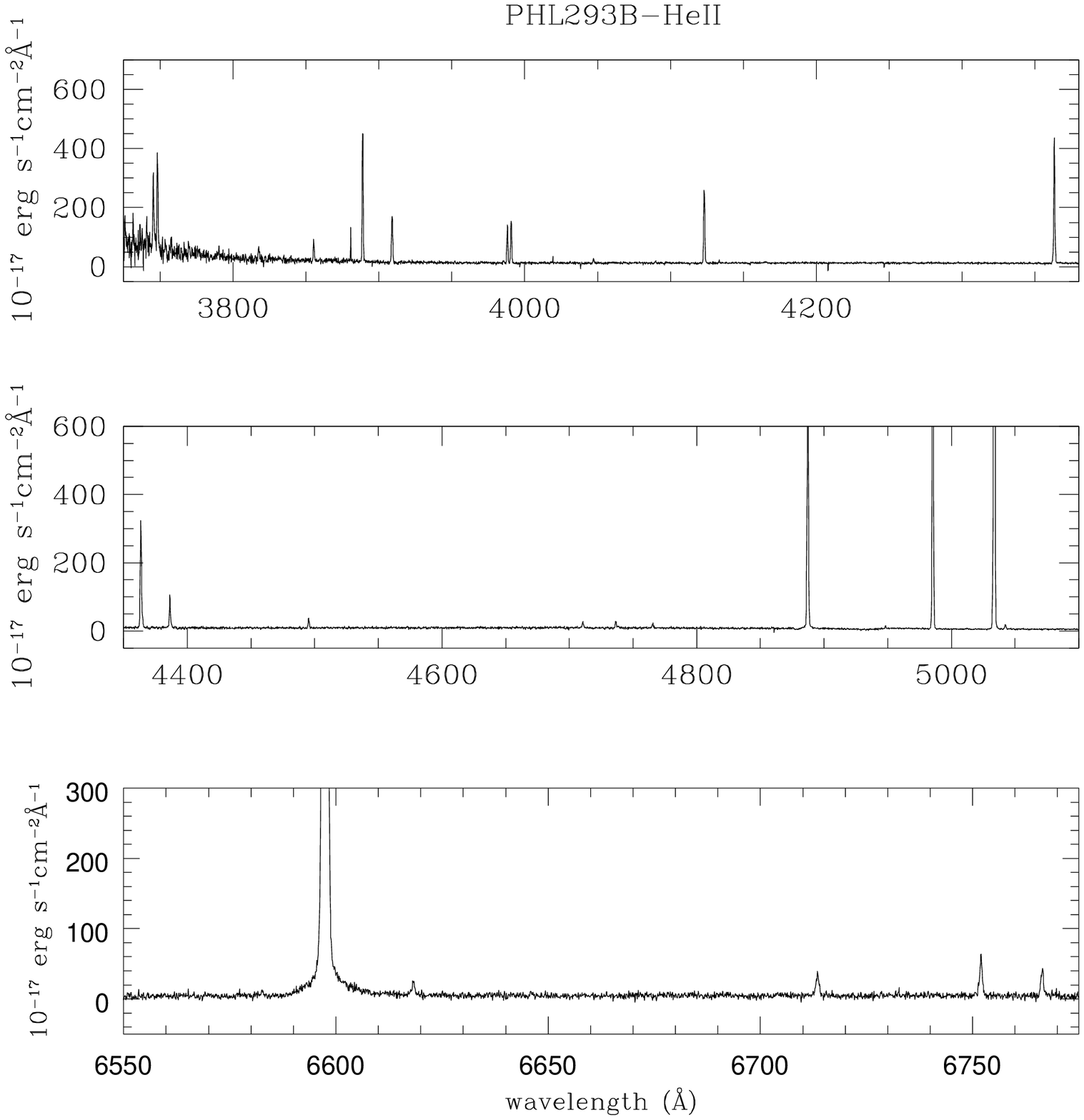}
\includegraphics[width=0.49\textwidth,clip]{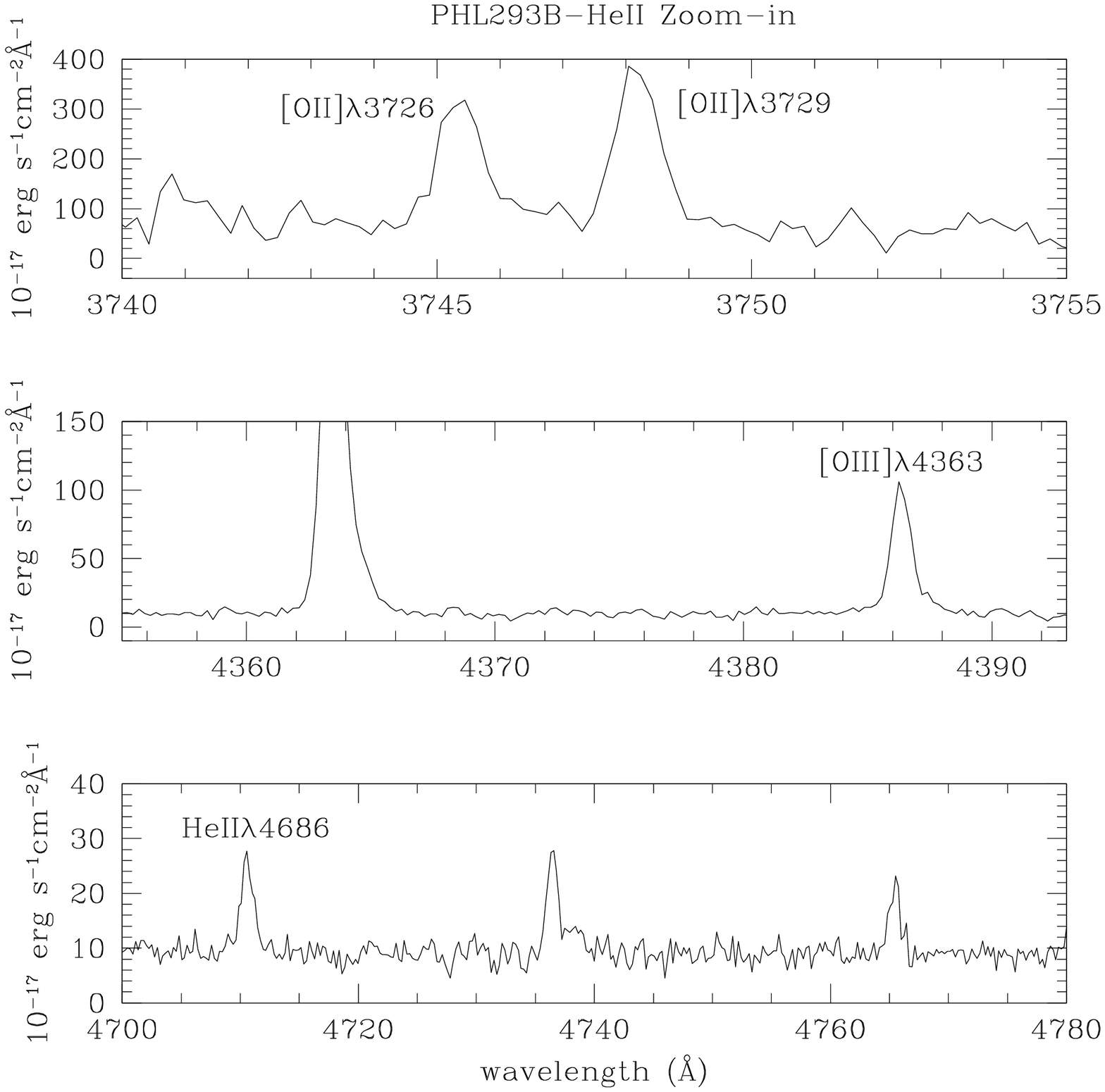}\\
\caption{Flux-calibrated spectra of the three regions defined in Section 5. The {\it y-axis} shows the flux in units of 10$^{-17}$ erg s$^{-1}$ cm$^{-2}$ \AA$^{-1}$.
{\it Left column, three first rows from top to bottom:} the VPH405-LR, VPH480-LR, and VPH665-HR spectra for the H$\alpha$-HSN region. {\it Right column, three first rows from top to bottom:} the VPH405-LR, VPH480-LR, and VPH665-HR spectra for the PHL~293B-integrated region. {\it Left column, the three last rows from top to bottom:} the VPH405-LR, VPH480-LR, and VPH665-HR spectra for the PHL~293B-HeII region. For illustrative purposes, in the case of the PHL~293B-HeII spectrum,  the {\it last three rows of the right column} display a zoomed-in view of the [OII] doublet, and of the wavelength ranges \mbox{$\sim$ 4355-4390 \AA~} and \mbox{$\sim$ 4700-4780 \AA~} showing the [OIII]$\lambda$4363 and nebular HeII$\lambda$4686 lines, respectively}
\label{spec_int} 
\end{figure*}

The 1D spectra mentioned above are presented in Fig.~\ref{spec_int}.
We derive the fluxes of the emission-lines and associated
uncertainties for these spectra using the same method as for
individual spaxels (see Section 3). We computed the logarithmic
reddening coefficient, C(H$\beta$), by performing a least square fit
to the ratio of the measured-to-theoretical Balmer decrements as a
function of the \citet[][]{mm72} galactic reddening law \citep[see
also][]{h08}. The uncertainty of the fit is adopted as the error in
C(H$\beta$). The narrow component of the four strongest Balmer
emission lines (H$\alpha$, H$\beta$, H$\gamma$, H$\delta$) have been
used. Intrinsic Balmer line ratios were taken from \citet[][]{OF06}
assuming case B recombination with electron temperature T$_{\rm
e}$=2$\times$10$^{4}$K: (H$\delta$/H$\beta$)$_{\rm theo}$=0.26,
(H$\gamma$/H$\beta$)$_{\rm theo}$=0.47, (H$\alpha$/H$\beta$)$_{\rm
theo}$=2.75.  Some issues were found concernig the C(H$\beta$)
computation. The blue grating covers the H$\delta$ and H$\gamma$, and
the green spectra include both H$\gamma$ and H$\beta$, while H$\alpha$
is measured in the red grating (see Section 2). We match the blue and
green spectra by using the H$\gamma$ line as reference which is
measurable in both. We could not connect the green and red spectra
since they share no lines. Thus, to minimize errors in the de-reddened
ratios between a certain emission line and H$\beta$, we always take
first its ratio in relation to the closest hydrogen emission line
(i.e. H$\delta$ in the case of [O{\sc ii}]$\lambda$3726, [O{\sc
ii}]$\lambda$3729 and [NeIII]3868$\lambda$; H$\alpha$ in the case of
[N{\sc ii}], HeI6678 and [S{\sc ii}]) and then we renormalize it using
the corresponding theoretical Balmer ratio \citep[e.g.,][]{EPM09,EPM11,K11}.

Table~\ref{table_regions} presents the relative fluxes of the
  de-reddened narrow emission lines measured from the integrated
  spectra; fluxes are normalized to the H$\beta$ flux = 1000.  We note
  that the values of C(H$\beta$) obtained here are in agreement with
  values derived for PHL~293B in the
  past \citep[e.g.,][]{I12,T14}. Also, the listed H$\alpha$/H$\beta$
  and H$\delta$/H$\beta$ ratios acceptably match their theoretical
  recombination values; the H$\gamma$/H$\beta$ ratio shown in
  Table~\ref{table_regions} makes use of the green-H$\gamma$ flux, and
  is about 10$\%$-15$\%$ smaller than the theoretical one. We verify
  that using the blue-H$\gamma$ flux instead, the ratio between the
  de-reddened and theoretical H$\gamma$/H$\beta$ lowers down to
  3$\%$-5$\%$. The line ratios uncertainties consider error flux
  measurements and the error in C(H$\beta$), but do not take
  systematic uncertainties, e.g., due to the blue-green match. We note
  that the effects of these uncertainties on the line ratios upon
  which oxygen abundance and $T_{\rm e}$[O{\sc iii}] estimates are
  based should be marginal since we obtain values in accord with other
  authors (see below). Since the [O{\sc ii}] lines are the most
  affected by extinction in our spectra, as a further check, we also
  corrected for reddening [O{\sc
  ii}]$\lambda$$\lambda$3726,3729/H$\beta$ using only the
  green-H$\gamma$-to-H$\beta$ ratio. We found that the variations in
  [O{\sc ii}]$\lambda$$\lambda$3726,3729/H$\beta$ are within the
  quoted uncertainties in Table~\ref{table_regions}.

For our three selected galaxy regions (PHL~293B integrated,
PHL~293B-HeII, H$\alpha$-HSN), we calculated physical properties and
oxygen abundances as explained in Section 4 for single spaxel
spectra. We calculated the nitrogen ionic abundance ratio,
N$^{+}$/H$^{+}$, from the PYNEB-based expression
from \cite[][]{EPM17}, using the [N{\sc ii}]$\lambda$6584 emission
line and assuming Te[N{\sc ii}] $\approx$ Te[O{\sc ii}]; we derived
the N/O ratio under the premise that N/O=N$^{+}$/O$^{+}$, based on the
similitude of the ionization potentials of the ions N$^{+}$ and
O$^{+}$.  Table~\ref{table_regions} also lists the values of
C(H$\beta$) and physical-chemical properties obtained for each
spectrum region.

By looking at Table~\ref{table_regions} and Fig.~\ref{bpt}, we find
  that the measurements of the integrated line-ratios [O{\sc
  iii}]$\lambda$5007/H$\beta$, [N{\sc ii}]$\lambda$6584/H$\alpha$,
  [S{\sc ii}]$\lambda$6717,6731/H$\alpha$ for the three selected
  regions are located below the demarcation lines in the BPT diagrams
  which implies an HII region-like ionization.

The comparison among the integrated electron temperature values from
Table~\ref{table_regions} shows similar $T_{\rm e}$[O{\sc iii}],
considering the uncertainties. Concerning the oxygen abundances, we
also find that the spectra of the regions PHL~293B-HeII,
H$\alpha$-HSN, and PHL~293B-integrated yield equivalent values within
the corresponding error bars. This is telling us that the metallicity
obtained from the PHL~293B-integrated spectrum matches the O/H from
the other two (physically smaller) selected zones. Therefore,
according to our MEGARA data, the gas metallicity of PHL~293B is not
only spatially homogenous (see previous section), but also independent
of the aperture applied. Here, we take the O/H abundance of the
integrated-spectrum [12+log (O/H)=7.64 $\pm$ 0.06 $\sim$ 8$\%$ solar
metallicity] as the representative metallicity of PHL~293B. This value
is consistent, within the errors, with those reported in previous
work \citep[e.g.,][]{K81,I07,P08,G09}.

Regarding the nitrogen abundance, we find that the H$\alpha$-HSN and
PHL~293B-HeII regions present similar N/O ratios within the
uncertainties (see Table~\ref{table_regions}). The N/O values derived here are in
agreement with the typical value of Log(N/O) $\approx$ -1.5 to -1.6
characteristic for the plateau in the 12+Log(O/H) vs. Log(N/O)
relation observed for low-metallicity
systems \citep[e.g.,][]{G90,thuan95,IT99,vansee,molla06,EPM11}. Moreover,
we confirm that the N/O ratios for these two regions match those
obtained in earlier studies of PHL~293B \citep[e.g.,][]{F80,I07}.

In Table~\ref{components} we list the observed fluxes and dispersion
of the broad and narrow emission components of H$\alpha$ and H$\beta$
lines for the three integrated spectra described in this section. The origin of the broad
emission and P Cygni-like features in the Balmer lines seen in the
spectra of PHL~293B has been debated for many years.
Discrepant scenarios involving a luminous blue variable star eruption,
an expanding supershell or a stationary wind driven by a young cluster
wind, and strongly-radiative stationary cluster wind have been
proposed \citep[e.g.,][]{IT09,T14,T15T}. \citet[][]{phl2}
review the previous interpretations for the nature of PHL~293B
including new 2019 Gemini data, and find a recent fading of
broad H$\alpha$ emission \citep[see also][]{phl1}; a broad to
narrow H$\alpha$ flux ratio (H$\alpha$ B/N) of 0.41 from 2001 SDSS
data and 0.10 from 2019 Gemini data are reported by \citet[]{phl2}.
Here we find H$\alpha$ B/N $\sim$ 0.10 for all
the three integrated regions indicating that the dissipation of the broad H$\alpha$
emission might have begun in 2017 when our observations were
performed (see Table~\ref{components}). However, while our data reveal P Cygni-like features in H$\beta$ and H$\alpha$  (see Figs.~\ref{pcyg_hb} and \ref{pcyg_ha}), P Cygni profile in H$\alpha$ is not visible in the 2019 Gemini spectra  
according to \citet[]{phl2}\footnote{Gemini spectra from \citet[]{phl2} cover from 5500 - 7500 \AA.}. A
long-lived Type IIn supernova (SN IIn) is proposed to be the most likely explanation
for the optical and spectral variability of PHL~293B by \citet[]{phl2}. However, the
lack of X-rays \citep[$\lesssim$ 3 $\times$ 10$^{38}$ erg s$^{-1}$; e.g.,][]{prest13,T14} in
PHL~293B remains the big challenge to the SN IIn scenario. Larger
timescales spectroscopic follow-up should be necessary to clarify the
variable spectral features of PHL~293B, but this is outside the scope
of our study.

\begin{table*}
\caption{De-reddened narrow emission line-fluxes relative to H$\beta$=1000 
and physical properties from three selected regions$^{a}$} 
\label{table_regions}
 \centering 
\begin{minipage}{15.0cm}
\centering
\begin{tabular}{lccc}
\hline\hline 
& \multicolumn{3}{c}{Selected regions} \\
Wavelength (\AA) & PHL~293B-Integrated$^{b}$   & H$\alpha$-HSN$^{c}$ & PHL~293B-HeII$^{d}$  \\  \hline
3726 [O~II]    & 585 $\pm$ 168     & 464  $\pm$ 46  & 296  $\pm$ 32     \\ 
3729 [O~II]    & 760  $\pm$ 118    & 544  $\pm$ 42  & 340  $\pm$ 27     \\
3868 [NeIII]   & 417  $\pm$ 30   & 418  $\pm$ 7  & 407  $\pm$ 6     \\
4100 H$\delta$ & 286  $\pm$ 16  & 277  $\pm$ 5  & 271  $\pm$ 4     \\
4340 H$\gamma$  & 390  $\pm$ 18   & 417  $\pm$ 10  & 412  $\pm$ 9     \\
4363 [O~III] & 84  $\pm$ 14   & 106  $\pm$ 6  & 113  $\pm$ 5     \\ 
4686 He~II   &   ---   &  20 $\pm$ 2  & 26  $\pm$ 2     \\ 
4714 [Ar~IV]  &   ---   & 21  $\pm$ 2   & 24  $\pm$ 3     \\
4740 [Ar~IV]  &   ---   & 14  $\pm$ 2  & 16  $\pm$ 2     \\
5007 [O~III]  & 3568  $\pm$ 38   & 4068  $\pm$ 44  & 4406  $\pm$ 32     \\
6563 H$\alpha$ (Narrow) & 2717  $\pm$ 55   & 2831  $\pm$ 39  & 2804  $\pm$ 28     \\
6584 [N~II]   &  ---   & 20  $\pm$ 3  & 16  $\pm$ 4     \\ 
6678 HeI    &   ---   & 24  $\pm$ 1  & 23  $\pm$  1   \\   
6717 [S~II]   & 82  $\pm$ 10   & 54  $\pm$ 2  &  36 $\pm$ 1    \\ 
6731 [S~II]   &  ---   & 36  $\pm$ 2  & 27  $\pm$  2   \\
c(H$\beta$)    & 0.11  $\pm$ 0.02     & 0.14 $\pm$ 0.01   & 0.30 $\pm$ 0.01  \\ 
\hline
F(H$\beta$) (erg s$^{-1}$ cm$^{-2}$) & 4.28$\times10^{-14}$ & 2.74$\times10^{-14}$ & 1.86$\times10^{-14}$   \\
F(HeII) (erg s$^{-1}$ cm$^{-2}$) & ---  & 5.41$\times10^{-16}$ & 4.76$\times10^{-16}$  \\
L(HeII)(erg s$^{-1}$)$^{e}$ & ---  & 3.45$\times10^{37}$ & 3.04$\times10^{37}$ \\
Q(HeII)(photon s$^{-1}$)$^{f}$ & ---  & 4.16$\times10^{49}$ & 3.66$\times10^{49}$ \\
\hline
log ([N~II]6584/H$\alpha$)  & ---  & -2.16  & -2.25     \\
log ([S~II]6717+6731/H$\alpha$)  & ---  & -1.50  & -1.66     \\
log ([O~III]5007/H$\beta$)  & 0.55 & 0.61  &  0.64    \\
$T_{\rm e}$([O~III]) (K) & 16,335 $\pm$ 1500   & 17,243 $\pm$ 548  & 17,089  $\pm$ 450   \\
$T_{\rm e}$([O~II])$^{g}$ (K) & 14,361  $\pm$ 1080    & 15,015  $\pm$ 394  & 14,904 $\pm$ 324  \\
12+log($O^{+}/H^{+}$)  & 7.08  $\pm$ 0.11    & 6.90 $\pm$ 0.04  & 6.71 $\pm$ 0.04   \\ 
12+log($O^{++}/H^{+}$)  & 7.50 $\pm$ 0.07     & 7.51  $\pm$ 0.02  & 7.55 $\pm$ 0.02  \\ 
12+log(O/H)$^{h}$  & 7.64 $\pm$ 0.06  & 7.60  $\pm$ 0.02  & 7.61 $\pm$ 0.02   \\ 
12+log($N^{+}/H^{+}$)  & ---    & 5.21 $\pm$ 0.07   &   5.12 $\pm$ 0.11   \\ 
log(N/O)  & ---    &  -1.68 $\pm$ 0.08   &  -1.59 $\pm$ 0.11    \\ 
\hline
\end{tabular}
\end{minipage}
\begin{flushleft}
(a) In all cases, the reddening correction for each line flux was performed relative to the closest Balmer recombination line (see text for details) \\
(b) PHL~293B-integrated spectrum obtained by co-adding all spaxels with H$\alpha$ S/N $>$ 3 \\
(c) Spectrum created by adding all spaxels with H$\alpha$ S/N $>$ 100 \\
(d) Spectrum obtained by summing all HeII-emitting spaxels \\
(e) HeII luminosity at the distance of 23.1 Mpc \\
(f) Number of ionizing photons shortward of the He$^{+}$ edge (see the text for details)\\
(g) $T_{\rm e}$[O{\sc ii}] = 0.72 $\times$ $T_{\rm e}$[O{\sc iii}] + 0.26 \citep{PI06}\\
(h) O/H = O$^{+}$/H$^{+}$ + O$^{2+}$/H$^{+}$\\
\end{flushleft}
\end{table*}

\begin{table*} 
\caption{Fit parameters of the broad and narrow hydrogen (H$\beta$, H$\alpha$) emission lines for the regions listed in Table\ref{table_regions}} 
\label{components} 
\centering 
\begin{minipage}{14.9cm} 
\centering 
\begin{tabular}{cccccc} 
\hline\hline  
Region$^{a}$ & Property & H$\beta$ Narrow & H$\beta$ Broad   & H$\alpha$ Narrow & H$\alpha$ Broad
\\ \hline 
\multirow{3}{*}{PHL~293B-Integrated}  & Flux$^{b}$   & 3296 $\pm$ 32  & ---  & 9720 $\pm$ 61 & 1005 $\pm$ 79   \\ 
                                      & $\sigma_{obs}$ (\AA)$^{c}$  & 0.50   & ---  & 0.46  & 2.3    \\ 
                                      & $\sigma_{obs}$ (Km/s)$^{d}$  & 31   & ---  & 21 & 105   \\  \hline
\multirow{3}{*}{H$\alpha$-HSN} & Flux$^{b}$  & 1978 $\pm$ 16  & 103 $\pm$ 21  & 6203 $\pm$ 15 & 589 $\pm$ 31   \\ 
                                         & $\sigma_{obs}$ (\AA)$^{c}$ & 0.49   & 2.22  & 0.46  & 3.17  \\  
                                         & $\sigma_{obs}$ (Km/s)$^{d}$  & 30    & 137  & 21  & 145   \\  \hline
\multirow{3}{*}{PHL~293B-HeII} & Flux$^{b}$ & 933 $\pm$ 5  & 65 $\pm$ 9   & 3247 $\pm$ 7   & 375 $\pm$ 19   \\ 
                                 & $\sigma_{obs}$ (\AA)$^{c}$ & 0.50    & 2   & 0.46   & 3.8  \\         
                                 & $\sigma_{obs}$ (Km/s)$^{d}$ & 31    & 123   & 21   & 174  \\ \hline
\hline 
\end{tabular} 
\end{minipage}
\begin{flushleft}
(a) Regions as defined in Table~\ref{table_regions}. \\
(b) Observed fluxes in units of 10$^{-17}$ erg s$^{-1}$ cm$^{-2}$. \\
(c) and (d) Observed $\sigma$ (= FWHM/2.35) in units of Angstrom and Km/s, respectively \\
\end{flushleft}
\end{table*} 

\section{The nebular HeII$\lambda$4686 in PHL~293B}

Photons with energy beyond 54 eV are needed to ionize He twice, so
HeII-emitting objects should host a relatively hard radiation field.
While nebular HeII emitters are atypical of nearby galaxies, they are
expected to be usual at high-z (z $\gtrsim$ 6) due to the predicted
harder UV-ionizing spectra at the lower metallicities typical in the
far-away Universe \citep[e.g.,][]{s15,DS16,se19}.  Next generation
telescopes (e.g., JWST, ELT) are expected to detect the rest-frame UV
of thousands of high-ionizing galaxies in the reionization
era. Therefore, studying the HeII-ionization in metal-poor local objects is
crucial to illuminate the properties of these reionization-epoch
systems.

It is to be noted that the fraction of HeII-emitting systems among
metal-poor objects tend to be larger than that for higher metallicity
galaxies observed in the local Universe \citep[e.g.,][]{K11,sb12}.
Ultra luminous X-ray binaries (ULXB), hot massive stars and shocks are
among the leading candidate sources discussed in the literature to
explain the nebular HeII excitation in nearby SF
galaxies \citep[e.g.,][]{g91,K11,sb12,dori,S20}. However, despite
observational and theoretical efforts, the origin of the He$^{+}$
ionization is far to be a settled matter in several
cases \citep[e.g.,][]{g91,K15,K18,plat19,kub19,zv20}. Current stellar
models keep failing to reproduce the total emergent flux beyond 54 eV,
specially in metal-poor galaxies \citep[e.g.,][]{K15,K18,se19}.

The existence of narrow HeII$\lambda$4686 emission in PHL~293B has
been noted before from long-slit
spectroscopy \citep[e.g.,][]{I07,P08,G09,I11}. Here, we produce the
first HeII$\lambda$4686 spectral map of PHL~293B using MEGARA (see
Fig.~\ref{line_maps}).  From our data, we checked that the FWHM of the
HeII$\lambda$4686 line matches that of other nebular emission lines
like the strong [O{\sc iii}]$\lambda$5007. The measured values of the
mean and standard deviation for the FWHM(HeII)/FWHM([O{\sc
iii}]$\lambda$5007) ratio are $\sim$ 1.10 and 0.10, respectively. The
narrow line profile for the HeII$\lambda$4686 emission and its spatial
extent are evidence of its nebular origin \citep[see also][]{sb12}.

PHL~293B was observed with the \emph{Chandra X-ray Observatory} on
2009 September for a total exposure time of 7.7 ks using the ACIS-S3
detector.  There is no detection of X-ray emission up to an upper
limit of $\sim$ 3$\times$10$^{38}$
erg~s$^{-1}$ \citep[][]{prest13,T14}. This indicates that X-ray
sources are unlikely to be the main responsible for the He~{\sc ii}
ionization in PHL~293B. On the other hand, the BPT line-ratios
measured both from the single HeII-emitting spaxels and integrated
spectra show values typical of HII region-like ionization (see
Fig.~\ref{bpt} and Table~\ref{table_regions}) indicating hot massive
stars as the dominant excitation source. This agrees
with \citet[]{phl2} who claim that the narrow emission gas in PHL~293B
is likely the HII region ionized primarily by stellar emission.
Wolf-Rayet (WR) emission bumps are not
detected in the spectra of PHL~293B. This means that different types of
hot stars other than WRs should be contributing to the HeII
excitation. This result agrees with the studies of the HeII-emitting
extremely metal-poor (XMP) galaxies IZw18 and SBS~0335-052E \citep[see][]{K15,K18}. A
detailed comparison of our observations to model predictions would be
needed to constrain the hot ionizing stellar population in PHL~293B,
but this exercise is beyond the scope of this paper.

For the PHL~293B-HeII spectrum (obtained by adding all the
HeII-emitting spaxels; see Fig.~\ref{line_maps} and
Section~\ref{int}), we computed the HeII ionizing photon flux, {\mbox
Q(HeII)$_{PHL~293B-HeII}$} = 3.66$\times$10$^{49}$ photons s$^{-1}$ (see
Table~\ref{table_regions}), from the corresponding reddening-corrected
luminosity L(HeII) using the relation {\mbox
Q(HeII)=L(HeII)/[j($\lambda$4686)/$\alpha_{B}$(HeII)]}
(assuming case B recombination, and $T_{\rm e}$([O{\sc
iii}]=2$\times$10$^{4}$ K; Osterbrock \& Ferland 2006). Applying the
same method for the H$\alpha$-HSN region, whose area includes
the PHL~293B-HeII region (see Fig.~\ref{line_maps} and
Section~\ref{int} for details) we find that the H$\alpha$-HSN
region produces Q(HeII)= 4.16$\times$10$^{49}$ photon s$^{-1}$ (see
Table~\ref{table_regions}). This is $\sim$ 14 $\%$ higher than
Q(HeII)$_{PHL~293B-HeII}$ which indicates that some small fraction of gas beyond the
PHL~293B-HeII region is also emitting He$^{+}$-ionizing photons. The
PHL~293B-HeII and H$\alpha$-HSN regions, together produce a
total Q(HeII) = 7.82 $\times$10$^{49}$ photons s$^{-1}$ which can be
taken as the HeII ionizing budget measured for PHL~293B. It is worth
noticing that the PHL~293B-integrated spectrum, created by summing
almost all the emission across the MEGARA FOV, does not show the HeII line
(see Table~\ref{table_regions}). In this regard one should bear in
mind that searches for reionization-era HeII-emitters, for which only
the total integrated spectra will be available, might be biased in
the sence shown here,i.e., that a non-detection of the HeII line does not necessarily
mean the intrinsic absence of HeII emission.

Using integral field spectroscopy (IFS), we also studied the spatial distribution of the nebular
HeII emission for the XMPs SBS~0335-052E and
IZw18 \citep[][]{K15,K16,K18}. When comparing the observed Q(HeII) for
different regions across SBS~0335-052E and IZw18, we find, for both
objects, that the highest absolute HeII flux and maximum Q(HeII)
values correspond to the integrated spectrum of the galaxy, contrary
to what we see in PHL~293B. This could suggest that the fraction of
HeII-ionizing hot stars, with respect to the total massive stellar content,
should be higher in SBS~0335-052E and IZw18 in comparison to the that
of PHL~293B, and that a higher amount of He$^{+}$-ionizing photons is
reaching larger distances from the central star clusters in both
SBS~0335-052E and IZw18. This might be related to the fact that,
although the three objects are very low-Z, the specific star
formation rate (sSFR) of SBS~0335-052E and IZw18
\citep[170 Gyr$^{-1}$  and 166 Gyr$^{-1}$, respectively;][]{sch16} is $>$ 20 times that of PHL~293B sSFR $\sim$ 6 Gyr$^{-1}$ \citep{F13}.  Of course, higher statistics is necessary to  make stronger
statements on which  properties can be dominant factors to determine the HeII emitting nature of a galaxy.

All the results described above testify the importance of IFS for this kind of
analysis, which allows us to collect all HeII emission, and
therefore deriving the absolute HeII ionization budget.

\section{Summary and conclusions}

We have analysed MEGARA observations of the nearby, very
metal-deficient galaxy PHL~293B. This kind of objects constitute
excellent laboratories for probing the conditions of galaxies in the
early universe. The data cover the optical wavelength range ($\sim$
3700-6800 \AA) within a field-of-view of $\sim$ 12.5 $\times$ 11.3
arcsec$^{2}$. MEGARA-IFU scans the entire spatial extent of the
PHL~293B main body providing us with a new 2D view of the ionized ISM
in this galaxy. Maps for the spatial distribution of relevant emission
lines, line-ratios and physical-chemical properties for the ionized
gas have been discussed. We were able to detect low intensity broad
components and P Cygni-like profiles in the Balmer lines in agreement
with previous work. We have checked that such components coincide
spatially with the brightest star-forming cluster of the galaxy.

The BPT-line ratios ([O{\sc iii}]$\lambda$5007/H$\beta$, [N{\sc
ii}]$\lambda$6584/H$\alpha$, [S{\sc
ii}]$\lambda\lambda$6717,6731/H$\alpha$) measured both from individual
spaxels and integrated spectrum regions agree with HII-like
ionization. We measured the [O{\sc iii}]$\lambda$4363 line flux over
the central parts of the galaxy covering an area of $\sim$ 0.12
kpc$^{2}$. For this zone, we measured O/H directly from the derived electron temperature T$_{e}$[OIII],
and we find no significant variations in oxygen abundance; most of
spaxels have 12+log(O/H) values spanning around $\approx$ 7.5--7.6. For
the first time, we derive the PHL~293B integrated spectrum by summing
the spaxels with H$\alpha$ S/N $>$ 3. We take the O/H abundance of the
PHL~293B integrated spectrum, 12+log (O/H)=7.64 $\pm$ 0.06 $\sim$
8$\%$ solar metallicity, as the representative metallicity of the
galaxy. Such value concurs with the ones on a spaxel-by-spaxel basis,
and it also matches with those found in the literature.

Here, we derive the first spectral map for the nebular
HeII$\lambda$4686 line and compute the HeII ionization budget in
PHL~293B. Our observations together with data from the literature
indicate that neither Wolf-Rayet stars nor X-ray binaries are the
main responsible for the HeII ionization in PHL~293B. This is in the
line of our studies on the two XMPs SBS~0335-052E and IZw18 based on
IFS. Additional IFS studies of large samples of very metal-deficient and nebular
HeII-emitters are needed to better understand the nature of these
objects.

\section*{Acknowledgements}

Based on observations made with the Gran Telescopio Canarias (GTC),
instaled in the Spanish Observatorio del Roque de los Muchachos of the
Instituto de Astrof\'{\i}sica de Canarias, in the island of La
Palma. This work is (partly) based on data obtained with MEGARA
instrument, funded by European Regional Development Funds (ERDF),
through Programa Operativo Canarias FEDER 2014-2020. We thank the
referee for a helpful report and thank M.A. Guerrero for useful
disussion. We acknowledge financial support from the Spanish Ministry
of Economy and Competitiveness under grant AYA2016-75808-R, which is
partly funded by the European Regional Development Fund, and from the
Excellence Network MagNet (AYA2017-90589-REDT). This work has been
partially funded by research project AYA2016-79724-C4-4-P from the
Spanish PNAYA. CK, JIP, JVM, SDP and EPM acknowledge financial
support from the State Agency for Research of the Spanish MCIU through
the "Center of Excellence Severo Ochoa" award to the Instituto de
Astrof\'{\i}sica de Andaluc\'{\i}a (SEV-2017-0709)

\section*{Data Availability}

The data underlying this article are part of the MEGARA commissioning
observations and are available in the article.

\bibliographystyle{mn2e}

\label{lastpage}
\end{document}